\documentclass[lettersize,journal]{IEEEtran} 
\usepackage{graphicx}
\usepackage{xspace}
\usepackage{xcolor}
\usepackage{listings}
\usepackage[T1]{fontenc}
\usepackage{caption}
\usepackage{array}
\usepackage{multirow}
\usepackage{url}
\usepackage{booktabs}
\usepackage{amsmath}
\usepackage{comment}
\usepackage{balance}
\usepackage[frozencache]{minted}

\newcommand{\totalCrawled}{438\xspace}
\newcommand{\totalInspected}{190\xspace}
\newcommand{\totalRemoved}{248\xspace}

\newcommand{\totalDebug}{129\xspace}
\newcommand{\totalImpl}{52\xspace}
\newcommand{\totalOpt}{9\xspace}
\newcommand{\totalConcepts}{19\xspace}

\newcommand{\todo} [1]{\textcolor{blue}{{\sf TODO}: #1}}
\newcommand{\codefont}[1]{\footnotesize{\texttt{#1}}\normalsize}

\newcolumntype{L}[1]{>{\raggedright\let\newline\\\arraybackslash\hspace{0pt}}m{#1}}
\newcolumntype{C}[1]{>{\centering\let\newline\\\arraybackslash\hspace{0pt}}m{#1}}
\newcolumntype{R}[1]{>{\raggedleft\let\newline\\\arraybackslash\hspace{0pt}}m{#1}}

\definecolor{codegreen}{rgb}{0,0.6,0}
\definecolor{codegray}{rgb}{0.5,0.5,0.5}
\definecolor{codepurple}{rgb}{0.58,0,0.82}
\definecolor{backcolour}{rgb}{0.95,0.95,0.92}

\def\BibTeX{{\rm B\kern-.05em{\sc i\kern-.025em b}\kern-.08em
    T\kern-.1667em\lower.7ex\hbox{E}\kern-.125emX}}

\begin{document}

%\title{Static Detection of Security API Misuses in Java Programs: Are We There Yet?}
%\title{Automatically Detecting Security-API Misuses in Java Programs: Are We There Yet?}
\title{Automation Configuration in Smart Home Systems: Challenges and Opportunities}
%\title{An Empirical Study on Automatic Detection of Security API Misuses in Java Programs}

\author{{Sheik Murad Hassan Anik, \and Xinghua Gao, \and Hao Zhong, \and Xiaoyin Wang, \and Na Meng}
\IEEEcompsocitemizethanks{\IEEEcompsocthanksitem S. Anik and N. Meng are with the Department of Computer Science, Virginia Tech, Blacksburg, VA 24060.
\protect\\
E-mail: \{murad, nm8247\}@vt.edu}
\IEEEcompsocitemizethanks{\IEEEcompsocthanksitem
X. Gao is with the Myers-Lawson School of Construction, Virginia Tech, Blacksburg, VA 24060. \protect\\
Email: xinghua@vt.edu
}
\IEEEcompsocitemizethanks{\IEEEcompsocthanksitem
H. Zhong is with the Department of Computer Science and Engineering, Shanghai Jiao Tong University, Shanghai, China. \protect\\
Email: zhonghao@sjtu.edu.cn
}
\IEEEcompsocitemizethanks{\IEEEcompsocthanksitem
X. Wang is with the Department of Computer Science, University of Texas at San Antonio, San Antonio, TX 78249. \protect\\
Email: xiaoyin.wang@utsa.edu
}
}

\maketitle

\begin{abstract}

As the innovation of smart devices and internet-of-things (IoT), smart homes have become prevalent. People tend to transform residences into smart homes by customizing off-the-shelf smart home platforms, instead of creating IoT systems from scratch. Among the alternatives, Home Assistant (HA) is one of the most popular platforms. It allows end-users (i.e., home residents) to smartify homes by (S1) integrating selected devices into the system, and (S2) creating YAML files to control those devices. Unfortunately, due to the diversity of devices and complexity of automatic configurations, many users have difficulty correctly creating YAML files. Consequently, their smart homes may not work as expected, causing frustration and concern in users.

This paper presents a novel study on issues of YAML-based automation configuration in smart homes (issues related to S2). We mined the online forum Home Assistant Community for discussion threads related to automation configuration. By manually inspecting \totalInspected threads, we revealed 3 categories of concerns: implementation, optimization, and debugging. Under each category, we classified discussions based on the issue locations and technical concepts involved.  Among debugging discussions, we further classified discussions based on users' resolution strategies; we also applied existing analysis tools to buggy YAML files, to assess the tool effectiveness. Our study reveals the common challenges faced by users and frequently applied resolution strategies. There are \totalDebug (68\%) examined issues concerning debugging, 
 but existing tools can detect at most 14 of the issues and fix none. It implies that existing tools provide limited assistance in automation configuration. 
  %support for users to correctly configure automation. 
  Our research sheds light on future directions in smart home development.

%Smart devices are getting more and more prominent in our day-to-day life. Each smart home system connects hundreds of smart devices. These devices are controlled by the automation configurations defined by developers. Writing the automation configuration for smart devices in smart home systems is a challenging task. A bug in automation configuration, i.e. cBug, can result in variable consequences, for example, improper behavior of the device or error in the whole system. In this work, we conducted the first empirical study on cBugs in Home Assistant, the most popular open-source smart home system. We investigated the root causes, impacts, and, resolution techniques of automation configuration bugs. The results of this study will help smart home developers better understand the automation configuration issues and open up new research directions for resolving automation configuration bugs in smart home systems.
\end{abstract}

\vspace{-.5em}
\section{Introduction}
% https://www.techtarget.com/iotagenda/definition/smart-home-or-building
A \textbf{smart home} is a residence that uses internet-connected devices to remotely monitor and manage appliances/systems.
According to Fortune Business Insights, the global Smart Home Market size is projected to reach USD 338.28 billion by 2030, at a Compound Annual Growth Rate (CAGR) of 20.1\% during the forecast period 2023–2030~\cite{fortune-smart-home}. 
As explained by researchers, the increasing number of internet users, surging disposable income of consumers within emerging economies, the growing significance of home monitoring in remote areas, and the increasing demand for low-carbon emission and energy-saving-oriented solutions are anticipated to drive the market competencies~\cite{smart-home-market}. 

% https://www.techtarget.com/iotagenda/definition/smart-home-or-building
%\textbf{Smart home technology}, also referred to as \textbf{home automation}, provides homeowners security, comfort, convenience, and energy efficiency by letting them control smart devices using a smart home app on their smartphone or another networked device.

%https://www.theverge.com/23751295/smart-home-platform-google-amazon-apple-samsung
A \textbf{smart home platform} is a software framework that controls and manages multiple devices from multiple manufacturers, usually through a smartphone or tablet app.
Various smart home platforms are available: some are commercial systems and close-source (e.g., 
Samsung SmartThings~\cite{smart_things});
% and Huawei HiLink \cite{huawei_hilink}); 
some are free and open-source (e.g., 
 openHAB~\cite{openhab_vs_ha}). 
 \emph{Among the alternatives, Home Assistant (HA) has become one of the most widely used platforms}~\cite{if-start-again,the-best,how-popular} mainly because it is free, open-source, and designed specially for local control as well as privacy.

Up till March 2024, HA has got over 332 thousand active installations; HA developers estimated the actual number of HA users (e.g., residents of HA-based smart homes) to be 3 times of this number (i.e., about 1 million)~\cite{haa}. 
 The widespread usage of HA motivated us to do research on HA-based smart homes, because (1) the data from these systems can represent many smart homes, and (2) our research findings will help shape the future of smart homes.

\begin{figure}
%\vspace{-1.em}
\centering
\includegraphics[width=.85\linewidth]{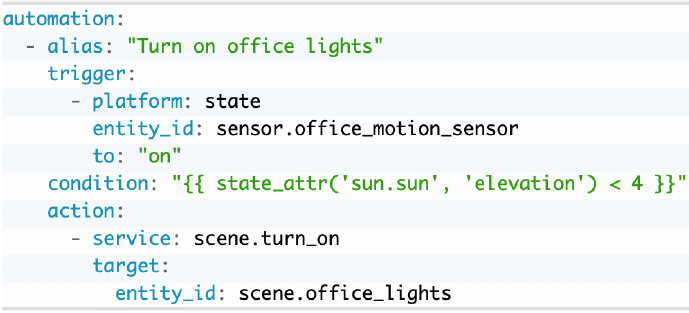}
\vspace{-.5em}
\caption{An exemplar YAML file for home automation~\cite{yaml-example}}
\label{fig:yaml-example}
\vspace{-1.5em}
\end{figure}

To create a smart home with HA, users need to (1) integrate components (e.g., devices) into the system and (2) create YAML files to automatically control those components. YAML~\cite{yaml} is a human-friendly data serialization language for all programming languages. With YAML, HA users can
define an automation rule by specifying a \textbf{trigger}, an \textbf{action}, and (optionally) a \textbf{condition}; such a rule
expresses that HA performs an action when a trigger event occurs and optionally the specified 
condition is met. For instance, the rule in Fig.~\ref{fig:yaml-example} means to turn on office lights, when someone moves (i.e., the sensor detects motion) and the outside is dark (i.e., the solar elevation angle is less than four degrees).

With some initial inspection of the online forum Home Assistant Community~\cite{hac}, we found lots of questions concerning automation configuration.
On the forum, over 10,000 questions were tagged with ``automation'' and over 1,000 questions were tagged with ``configuration''~\cite{tags}.
 This may be because many HA users have no programming background 
 %existing YAML editors provide limited coding assistance, the components-to-integrate are very diverse as well as hard to configure, 
 and automatic control is challenging. As a result, misconfigured smart homes may misbehave~\cite{restart-automations}, waste energy~\cite{house-thermostat}, frustrate users~\cite{going-crazy}, or jeopardize home safety.

Inspired by our initial observations, we conducted a novel empirical study on automation configuration issues in HA-based smart homes, to understand the technical challenges and identify research opportunities. Specifically, we crawled the discussion threads of ``Configuration'' during 03/2022--08/2022 using tag ``automation'', and retrieved \totalCrawled potentially relevant threads. Then we manually analyzed all threads to identify the root cause and resolution strategy of each issue under discussion, and filtered out \totalRemoved threads because they lack necessary information. We explored the following research questions (RQs) and observed interesting phenomena: 
%to understand the challenges in smart-home development and opportunities in future tool support. Therefore, we crawled 

%\vspace{-.5em}
\begin{itemize}
\item \textbf{RQ1:} \emph{What challenges do users face when configuring automation?} Among the \totalInspected threads, \totalDebug, \totalImpl, and \totalOpt discussions separately focus on debugging, implementation, and optimization. It means that users get stuck with debugging more often than other issues.
%43, 23, 73, and 52 threads separately focus on issues in trigger, condition, action, and the entire automation.
We also observed significant concept commonality/similarity among threads (e.g., data specification for matching), which implies users' strong need for help or tool support in implementing, debugging, or optimizing some features. %This implies that action is harder to configure than the other two components; many issues require for simultaneous revision of multiple components (e.g., trigger + condition) instead of singles.

%226 cBugs discussed in the \totalInspected threads, 72 and 71 cBugs are separately concerning entities and logic. It means that users or developers often get confused about (1) how to use integrated entities or components and (2) how to correctly define the control logic.
\item \textbf{RQ2:} \emph{How do users address challenges in automation configuration?} Users frequently applied eight strategies to address common challenges. 
%five strategies 
%frequently applied. 
Two of the strategies correct formats (i.e., quotes and indentation); two conditionally call various services; 
one replaces the trigger type; one correctly accesses or calculates data; one correctly specifies data for matching; one handles a group/list of same-typed entities. All strategies imply desired tool support. 
% (1) to correct misspelling, (2) to correct the format of quoted string values, (3) to correct variable declarations by moving them out of quoted strings, (4) to call method \codefont{delay()} for a better timing control, and (5) to call \codefont{timeDelta()} for the time difference between events in an automation.  
\item \textbf{RQ3:} \emph{How effectively do existing tools detect or fix buggy YAML files?} 
%Because HA-based automation configurations are defined as YAML files, 
We searched online for all existing tools that can validate YAML files. By applying the 6 publicly available tools to \totalInspected buggy files, we observed all tools to report only 16--20 files/snippets as invalid ones, and 1--5 of these reports are false positives. All tools achieved high precision, very low recall, and very low F score when detecting bugs; their error-reporting mechanism is poor. 
% they output poor error messages in many cases. 
No tool fixes bugs.
\end{itemize}
%\vspace{-1em}

%Although several studies characterized bugs in smart devices
In this paper, we made the following research contributions:

\begin{itemize}
\item We conducted a novel and comprehensive empirical study to characterize real-world issues of automation configuration in HA-based smart homes. No prior work did that.%characterized such issues. 
%this kind of real-world issues in IoT systems in such a systematic and comprehensive way.
\item We revealed the common root causes and recurring resolution strategies for 
HA-related automation issues; 
many of our findings are revealed for the first time.
%none of our findings were mentioned by any prior study on IoT systems.
\item We novelly applied 6 publicly available tools to \totalDebug buggy YAML files/snippets, and surprisingly found all tools ineffective in detecting or fixing bugs.
\item We open-sourced our dataset and tool-application results, to facilitate future research in related areas.
%facilitate future empirical studies and tool development in related areas.
%. Our data will facilitate future empirical studies and tool development in related areas.
\end{itemize}
HA is one of the most widely used smart home platforms, and \ commonly shares the automation configuration mechanism (if-this-then-that rules) with many other platforms (e.g., IFTTT~\cite{ifttt} and Samsung SmartThings~\cite{smart_things}). Thus, our study %characterization study of automation issues on HA-based systems 
will shed light on future research in smart homes, help with end-user IoT programming, and enlighten potential ways of improving smart-home quality. Our dataset is available at \url{https://figshare.com/s/7aa8ea9f4af98c371114}

\section{Background}
This section introduces terms relevant to Home Assistant. 

\begin{figure}
\centering
\vspace{-1.5em}
\includegraphics[width=.95\linewidth]{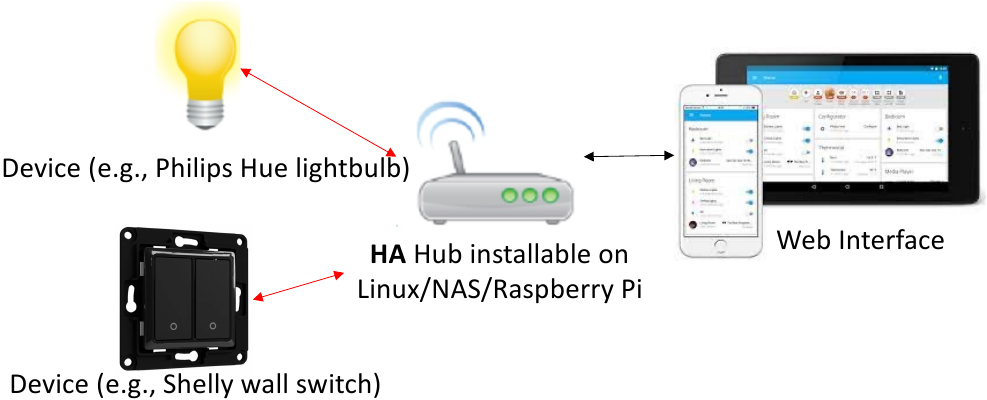}
\vspace{-.5em}
\caption{The consumer pattern in Home Assistant}
\label{fig:ha-pattern}
\vspace{-1.5em}
\end{figure}

\textbf{HA} can be installed on various devices, 
%is an open-source home automation hub that can be installed on a variety of devices---
from full Linux systems to some network-attached storage (NAS) environment or even a Raspberry Pi. As shown in Fig.~\ref{fig:ha-pattern}, users can access HA through a dashboard or web-based user interface by using companion apps for smart phones, or by using web browsers for tablets and PCs. 
% or by voice commands via a supported virtual assistant such as Google Assistant or Amazon Alexa. 
Once the HA software is installed, it acts as a hub---a central control system for home automation. Without any mandatory dependence on vendor-specific cloud services, devices, or mobile apps, 
 the HA hub can have local control of the IoT devices, software, applications, and services that are supported by modular integration components. 
\textbf{{Home Assistant Community (HAC)}}~\cite{hac} is an online forum dedicated for HA developers/users to discuss and resolve issues in smart home systems. HAC contains discussion threads. Each thread has one question post, and zero or more answer posts; at most one answer in each thread is the accepted answer.
 
 \textbf{Integrations} are pieces of software that allow HA to connect to other software and platforms. 
For example, a product by Philips called Hue smart light (see Fig.~\ref{fig:ha-pattern}) can be included into a smart home via the Philips Hue integration, which integration allows HA to talk to the hardware controller Hue Bridge, so that any HA-compatible physical devices connected to Hue Bridge appear in HA as logical devices (i.e, virtual objects) and can be controlled by the hub.

%\vspace{-.5em}
%\paragraph{2.2 Entities \& Related} 
\textbf{Entities} are the basic building blocks to hold data in HA. An entity represents a sensor, actor, or function in HA, which can monitor physical properties or control other entities. %An entity will write their current state to the state machine for other entities/templates to access. 
Each entity has a \textbf{state} to hold information of interest (e.g., whether a light is on or off); an entity's state only holds one value at a time. Entities can store \textbf{attributes} related to its state, such as the brightness of a turned-on light. 
%\textbf{State} is a current representation of the entity. 
\textbf{Sensors} return information about an object, such as the level of water in a tank.
%Entities are used to monitor physical properties or to control other entities. An entity is usually part of a device or a service. 
%\textbf{Service} carries out one specific task, such as turning on the light in the living room. A service has targets and data; it can be called by actions, a dashboard, or via voice command. 
\textbf{Devices} are logical groups for entities. A device may represent a physical device with one or more sensors; the sensors are entities associated with the device. %For example, a motion sensor is represented as a device. It may provide motion detection, temperature, and light levels as entities. 

\textbf{Automation} is a set of repeatable actions that can be run automatically. \textbf{Automation Configuration} is about defining YAML files to specify automations, after HA is installed and all components are integrated.
In a YAML file, an automation rule has three key segments: 
%Typically in a YAML file, users define an automation rule to include three key components:% (i) trigger, (ii) condition, and (iii) action.
%Listing \ref{lis:automation_def} intuitively explains the semantic meaning of the basic automation configuration shown in Fig.~\ref{fig:yaml-example}. 

\textbf{(i) Trigger} describes what starts an automation~\cite{trigger}. An automation can be triggered by an event (i.e., signal emitted when something happens), certain entity state (e.g., when a light is on), or a given time. Multiple triggers can be specified simultaneously for one automation.  When any of the automation's triggers becomes true (i.e, trigger fires), HA will validate the conditions if any, and call the action. For the example in Fig.~\ref{fig:yaml-example}, the trigger is an event: when the state of a motion sensor is changed to ``\codefont{on}''.
 
\textbf{(ii) Condition} is optional; it describes the \textbf{predicates} (i.e., tests) that must be met before actions get run~\cite{yaml-example}. After a trigger occurs, all conditions are checked. If any of them returns false, the action is not run. 
% the automation stops executing. %Conditions look similar to triggers, but they are different: a trigger will look at events happening in the system, while a condition only looks at the current system status.
%A trigger can observe that a switch is being turned on; a condition can only see if a switch is currently on or off. 
The condition in Fig.~\ref{fig:yaml-example} tests whether the solar elevation angle is less than four degrees.  

\textbf{(iii) Action} describes what is executed when a rule fires~\cite{action}. 
It can interact with anything via services or events.
\textbf{Service} carries out a task, such as turning on the light in the living room. A service can have a target and data.
%; it can be called by actions, a dashboard, or via voice command. 
 For instance, entity \codefont{scene.office\_lights} in Fig.~\ref{fig:yaml-example} is a \textbf{scene} that prescribes a series of actions, with each action setting an entity's state. The defined action interacts with this scene by calling service \codefont{scene.turn\_on}, to turn on lights and set their states as prescribed by the scene. 
 
\textbf{Scripts} are also repeatable actions, similar to automations. The difference is that scripts do not have triggers: scripts cannot automatically run unless they are used in an automation.

\textbf{Templates} are used for formatting outgoing messages to present to users or processing incoming data from entities. They are expressed with Jinja2~\cite{Jinja}---a general-purpose templating language. 
% powered by the \textbf{Jinja2}~\cite{Jinja} templating engine; Jinja2 is a general-purpose templating language.
% special placeholders in Jinja2 templates allow writing code similar to Python syntax; then the templates are passed data to render the final document. %In other words, users can use Jinja2 to define HA variables and data-handling logic.
%process input, or render output. 
For instance, \codefont{"\{\{state\_attr('sun.sun', 'elevation') < 4\}\}"} in Fig.~\ref{fig:yaml-example} is a template, which gets the solar elevation degree from entity \codefont{sun.sun} and compares that value with 4 to define a predicate for condition checking. 
%Fig.~\ref{fig:template-example} shows an exemplar template in the message data of a notification service. This script automatically generates notification messages, depending on the state value of device ``\codefont{device\_tracker.paulus}''. Namely, if the state is ``home'', the message is ``Ha, Paulus is home''; otherwise, the message is ``Paulus is at (another location state)''. %As shown in this example, 
%Templating allows automations to behave differently given different inputs. 
Templates enable smart homes to flexibly respond to varying entity states. To define templates, people must surround single-line
templates with double quotes (") or single quotes (’). When defining multi-line templates, people must use \codefont{\{\% \%\}} to enclose 
the program structures (e.g., loop) in templates.

\section{Methodology}

To understand the challenges and opportunities in home automation, we explored the following three research questions:

\begin{itemize}
\item\textbf{RQ1:} \emph{What challenges do users face when configuring automations?} What kind of automation-related questions do users ask? 
Are there questions frequently asked? 
What are the root causes of the frequently asked questions? %bugs (cBugs) do people encounter in HA-based smart home development?

\item\textbf{RQ2:} \emph{How do experts/users address challenges related to automation configuration?} Namely, when people ask similar or identical questions, is there any answer repetitively suggested? 
Do answers present resolution strategies? 
%What are the resolution strategies for frequently asked questions?

% what are the fixing strategies people often apply to fix cBugs?
\item\textbf{RQ3:} \emph{How effectively do existing tools detect or fix buggy YAML files?}
When users have difficulty debugging erroneous YAML files, can existing tools reveal or fix those errors? 
%What is the gap between (1) the complexity of bug detection/fixing, and (2) the capability of current tools?

% Namely, what is the gap between (1) the complexity of cBug detection as well as fixing, and (2) the capability of existing tool support?
\end{itemize}

This section first introduces our procedure of data collection, and then explains our method of exploring RQs.

\subsection{Data Collection}
The program data related to HA-based smart home systems can be majorly found on two websites: GitHub~\cite{github} and HAC~\cite{hac}. 
We decided to crawl HAC for two reasons. First, 
%we focus on bugs and fixes related to automation configurations in HA-based smart homes. 
HAC organizes discussion threads based on categories and tags.  
%every week, there are more than 300 automation configuration questions posted. 
Such an organization 
%of discussion threads 
%and the consistent growth of relevant discussions 
enables us to quickly locate automation configuration-related program data.
Second, many questions on HAC are resolved, with some resolutions well explained and finally accepted by askers. 
Such a high availability of technical solutions and rationale explanation enables us to characterize questions and answers with high rigor.
%although not every technical question is immediately resolved, most of the questions are solved eventually. 

Specifically, we manually went through the discussion threads under category ``Configuration'' of HAC, to locate candidates tagged with ``automation'' during 03/01/2022 - 08/31/2022. 
That six-month period was chosen because we collected data in September 2022. 
As HAC ranked threads based on timestamps of the latest activity on each thread (e.g., question posting or answer updating), we successfully retrieved a dataset of \totalCrawled candidate discussion threads purely based on the HAC ranking and timestamp range.
Next, we manually inspected each thread to record (1) ID of the question post, (2) the URL, (3) the original YAML file provided by the asker, and (4) the suggested YAML file mentioned, referred to, or implied by the accepted answer. In this process, we filtered out a thread if 

\begin{itemize}
\item[(i)] the question asker does not present a YAML file, 
\item[(ii)] no answer was explicitly marked as accepted by the asker,
\item[(iii)] the accepted answer does not suggest or lead to any concrete version of YAML file, or 
\item[(iv)] the discussion is too confusing for us to understand. 
\end{itemize}

We defined the four filters mentioned above, to ensure the high quality of our data analysis results. 
With more details, based on our experience so far, when people asked questions at HAC, they sometimes failed to provide YAML files to clearly show the programming context or automation scenarios. Consequently, analyzing such questions can be very challenging and time-consuming, as it is hard to tell what is the major issue and what types of resolutions askers inquiry for. To avoid such confusion, we introduced filter (i) to guarantee that each covered question post provides sufficient details on the programming context, including but not limited to askers' automation needs, the attempts they made, the technical issues they face, and their specific requests for help.
When processing each thread, we manually inspected all posts involved in the discussion. As long as the asker provided their YAML files in any of the posts, we kept those threads.

Filter (ii) ensures that there is a correct answer for us to refer to, when characterizing people's concern and resolution for each covered thread.
Filter (iii) makes sure that the suggested resolution is clear and concrete. Namely, we do not have to speculate on the correct version of YAML files, and thus will not commit mistakes in speculation.
In reality, however, an asker might accept an answer that provides no concrete YAML files, while the concrete version was mentioned in a different post of the same thread.
To tolerate such inaccuracy in askers' labeling of accepted answers, 
we read all posts in each thread. As long as a concrete YAML file is suggested by an accepted answer or any answer related to that one,
% an accepted answer included a concrete YAML file, we kept the thread. 
%the answer did not include a concrete YAML file but the other posts %related to that answer suggested a correct one, 
we counted in the thread.
% and treated that YAML file as the reference answer. 
In this way, we made our dataset as representative as possible. 
Filter (iv) ensures that we have high confidence in our interpretation of  threads.
After applying all filters mentioned above, 
%After removing threads that do not satisfy the above requirements, 
we kept \totalInspected threads for further analysis. 
%Depending on the location of bugs in the discussed YAML files, we classified these threads into four categories: 

\begin{figure*}
\includegraphics[width=\linewidth]{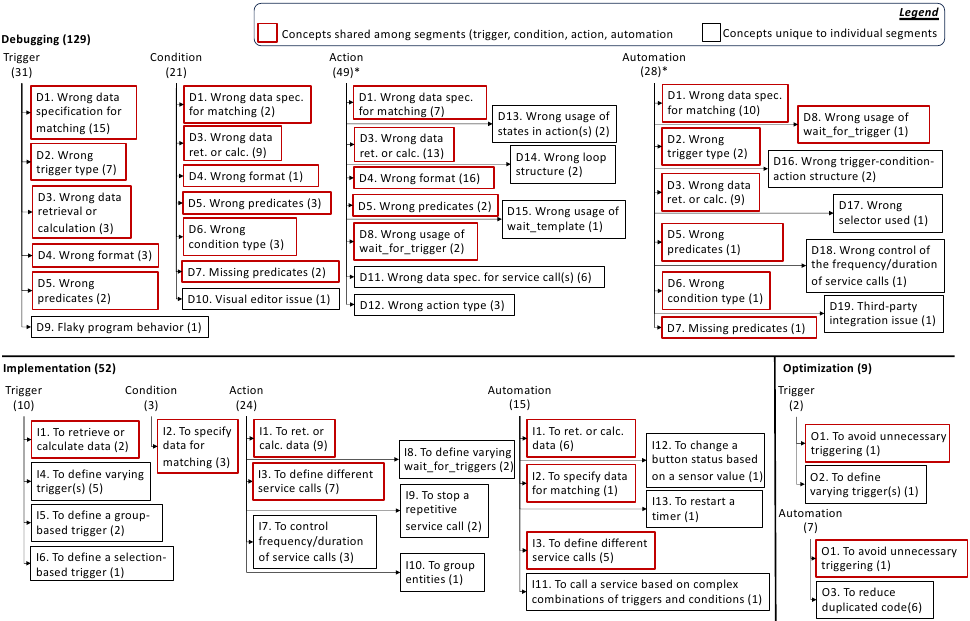}\vspace{-1.em}
\caption{The taxonomy of \totalInspected threads, where * means some issues belong to more than one category}\label{fig:taxonomy}
\vspace{-1.2em}
\end{figure*}

\subsection{Data Analysis}For RQ1, we took an open-coding approach to classify threads, because we had no prior knowledge of people's concerns on automation configuration. Specifically, two authors manually inspected all threads to identify keywords/phrases, which characterize each thread in terms of the (1) issue type, (2) automation component under discussion, and (3) involved technical concepts. Then they separately defined an initial taxonomy and categories by clustering recurring or similar keywords/phrases. 
Next, they held a meeting to compare and discuss their initial results to refine the taxonomy and improve thread classification. 
Using the classification labels they agreed upon, author A rechecked all threads to label categories and to reveal overlooked or wrong categories. Afterwards, author B examined A's labels for all threads; whenever he disagreed upon any labels, he had discussions with author A until reaching a consensus. This procedure sometimes involved multiple iterations of thread labeling, and lasted until both authors agreed upon all labels.

For RQ2, we also took an open-coding approach to identify frequently adopted resolution strategies. With more details, inside each category, author A summarized the resolution suggestion for each issue-under-discussion, and then clustered issues if their resolutions are identical or similar in certain aspect. In this procedure, A also defined resolution strategies based on the common summaries. 
%the summarized keywords or phrases. 
Next, author B manually inspected all strategies as well as related threads, and initiated discussion with A for any disagreement. The discussion lasted until all disagreements were addressed.

For RQ3, we first defined a ground-truth dataset of buggy YAML files/snippets, according to the classification results in RQ1. 
In this dataset, we also included the fixes suggested by accepted answers. Next, we applied all state-of-the-art YAML validation tools to the dataset, to study how effectively existing tools detect and fix bugs.
% those bugs, and (2) how effectively these tools fix bugs. 
%Based on tools' outputs and our assessment of those outputs, we compared tools. 
%the (i) the number of bugs they report, (ii) the accuracy of bug reports, (3) the number of fixes they suggest, and (4) the accuracy of bug-fixing suggestions.

\section{Experiment Results}\label{sec:result}

This section presents and explains our results for RQ1--RQ3.

\subsection{RQ1: Challenges in Handling Automation Configuration}\label{sec:rq1}

Fig.~\ref{fig:taxonomy} shows our thread classification based on issue types, issue locations, and involved technical concepts. 

%\vspace{-.5em}
\subsubsection{Classification Based on Issue Types}
We observed three types of technical issues: (1) implementation, (2) debugging, and (3) optimization. \textbf{Implementation} means that people described their program context and automation requirements; they asked for 
implementation to fit the context and satisfy those requirements.
\textbf{Debugging} means that users provide erroneous automations, requesting to diagnose the root cause and fix bugs. Compared with implementation, debugging questions often include askers' unsuccessful trials (i.e., buggy YAML files) to implement certain requirements.
 \textbf{Optimization} means that people present the correct YAML files satisfying certain requirements, express their optimization goals, and ask for refined YAML files.
 %for help in realizing automation refinement. 
 
Our dataset includes \totalDebug debugging issues, \totalImpl implementation issues, and \totalOpt optimization issues. It means that most askers had difficulty debugging YAML files, while fewer askers were bothered by implementation and optimization issues. This may be because automation configurations have relatively simple control logic. It is not quite difficult for people to get started and have initial implementation done. However, it is harder for people to successfully debug  initial implementation, in order to deliver high-quality automation configurations. Consequently, even fewer people bother to optimize correct configurations for refinement.

\vspace{.2em}
\noindent\begin{tabular}{|p{8.5cm}|}
	\hline
	\textbf{Finding 1:} \emph{68\% (129/190) of the examined issues are about debugging, implying the significant challenge of addressing buggy YAML files}.\\ \hline
\end{tabular}
\vspace{.2em}

\subsubsection{Classification Based on Segments Involved}
Under each issue category, we classified threads based on where issues occurred.
Namely, if an issue resided in purely the trigger, condition, or action segment of an automation rule, we labeled it accordingly. Otherwise, if an issue resided in multiple segments or we could not clearly tell which specific segment an issue was about, then we labeled the issue with ``Automation''. 

As shown in Fig.~\ref{fig:taxonomy}, for both debugging and implementation issues, ``Action'' is the largest among four categories. This implies that actions are harder to develop than triggers and conditions. One reason is that HA defines the script syntax~\cite{script-syntax}, which allows people to define complex control structures (e.g., if-then and loop), or special behaviors in actions (e.g., \codefont{wait\_for\_trigger} to wait for a trigger before calling a service). %While the syntax allows people to flexibly define actions and control actuators,
The syntax poses challenges for people to understand and properly use all syntactic components.
Additionally, there are more threads on triggers than conditions. %Specifically, among debugging issues, we observed 31 and 21 issues separately related to triggers and conditions. Under the implementation category, there are 10 and 2 issues separately about triggers and conditions. Regarding optimization, two issues focus on triggers and none is about condition. 
It implies that triggers are harder to develop, probably because there are more alternative ways of defining triggers, and more delicate constraints on trigger definition.
Finally, %Among all three types of issues, we observed the wide existence of automation issues. 
automation issues widely exist in all three categories. 
%In particular, among the nine optimization issues, seven are about whole-automation optimization. Such a wide existence of automation-related issues indicates that
It means that users have difficulty in developing multiple segments for consistency, or in creating code that can exist in any of the segments.

\vspace{.2em}
\noindent\begin{tabular}{|p{8.5cm}|}
	\hline
	\textbf{Finding 2:} \emph{There are 73, 50, 43, and 24 issues separately about action, automation, trigger, and condition, implying that automation and action segments are more challenging to create or improve}.\\ \hline
\end{tabular}
\vspace{.2em}

\subsubsection{Classification Based on Technical Concepts}
We further classified issues based on the technical concepts involved. In this procedure, whenever we observed similar issues in different segments (i.e., trigger, condition, or action), we used the same terms to capture the similarity, and to  identify segment-specific concepts as well as segment-agnostic ones. In Fig.~\ref{fig:taxonomy}, 
\textbf{the \totalDebug debugging issues cover \totalConcepts unique concepts; 8 concepts are shared among segments (D1--D8)}. 

\emph{D1. Wrong data specification for matching} means that data (e.g., state or attribute) was wrongly provided for predicate-value matching in trigger, condition, action, or the whole automation, making the automation run abnormally. 
For instance, lines 2--6 in Listing~\ref{lst:match} show a buggy snippet from a debugging thread~\cite{on}. This snippet defines a state trigger, to start an automation when an entity \codefont{input\_boolean.ecobee\_fan\_on\_off} changes from the \codefont{'off'} state to \codefont{'on'}. However, as both states are misspelled, HA is case-sensitive and does not recognize those states. Consequently, the automation is never started because the prescribed state transition never happens. 

\begin{listing}[h]
  \caption{Two YAML snippets to show the common challenge of ``data specification for matching''~\cite{on,not-to}}
    \vspace{-.5em}\label{lst:match}
  \begin{minted}[
    frame=single,
    fontsize=\scriptsize,
    linenos,
    numbersep=2pt,
    breaklines]{yaml} 
# Bug: Wrong states are specified for the to: and from: options    
trigger:
- platform: state
  entity_id: input_boolean.ecobee_fan_on_off
  to: 'On'     # Fix: 'On' should be 'on'
  from: 'Off' # Fix: 'Off' should be 'off'
# Implementation: To properly specify an unwanted target state with the not_to: option
trigger:
- platform: state
entity_id:
- sensor.pc1_printjob
- sensor.pc2_printjob
- sensor.pc3_printjob
- sensor.pc4_printjob
not_to: 'unknown'
\end{minted}
  \vspace{-.5em}
\end{listing}

\begin{listing}
\vspace{-.5em}
  \caption{A bug and fix related to D2~\cite{template-in-trigger-question}}\label{lst:trigger}
%    \vspace{-.5em}
  \begin{minted}[
    frame=single,
    fontsize=\scriptsize,
    linenos,
    numbersep=2pt,
    breaklines]{yaml} 
# Bug: A template is involved in a device trigger definition   
trigger:
  - type: no_motion
    platform: device
    device_id: e0954ea41d7a6da69baeff2e9558ed13
    entity_id: binary_sensor.v4b01_motion
    domain: binary_sensor
    id: '1'
    for:
      hours: 0
      minutes: ''{{states('input_number.timeout_offices') |int(0) }}'' 
      seconds: 0
# Fix: Replace the device trigger with a state trigger to support template usage  
trigger:
  - platform: state
    entity_id: binary_sensor.v4b01_motion
    id: '1'
    to: 'off'
    for:
      minutes: "{{ states('input_number.timeout_offices') | int(0) }}"   
  \end{minted}
  \vspace{-.5em}
\end{listing}

\emph{D2. Wrong trigger type:} There are multiple types of triggers (e.g., state trigger and device trigger) usable to specify the triggering logic of an automation. However, when users wrongly choose a trigger type, the specified logic does not work because the chosen type does not support that logic. For instance, the buggy version in Listing~\ref{lst:trigger} defines a device trigger, to fire when sensor \codefont{binary\_sensor.v4b01\_motion} detects no motion for the number of minutes users specified via \codefont{input\_number.timeout\_offices}, a numeric input box in GUI.
However, device triggers do not support templating (see line 11), so the buggy version fails. 

\begin{listing}
  \caption{Two YAML snippets to show the common challenge of ``data retrieval or calculation''~\cite{impl-state,debug-state}}  %\vspace{-.5em}
  \label{lst:retrieval}%\vspace{-.5em}
  \begin{minted}[
    frame=single,
    fontsize=\scriptsize,
    linenos,
    numbersep=2pt,
    breaklines]{yaml} 
# Bug: Wrong way of retrieving the sensor state value    
action:
- service: notify.mobile_app_huawei_p20
  data:
    title: Temperature Warning
    message: 'Temperature is: {{ sensor.temperatur_gefrierschrank }}' 
# Fix: sensor.temperatur_gefrierschrank =>   states("sensor.temperatur_gefrierschrank") 
# Implementation: To include sensor value into the notification
action:
  - device_id: c593fadfd2c2d134bc507567b588b2ae
    domain: mobile_app
    type: notify
    message: Energy production ended. Energy produced today: {{states('sensor.solaredge_current_power')}}    
\end{minted}
\vspace{-.5em}
\end{listing}

\emph{D3. Wrong data retrieval or calculation} means that some data used in an automation is wrongly accessed or calculated, causing automation fail to run normally. For instance, lines 2--6 in Listing~\ref{lst:retrieval} show a buggy snippet from a debugging thread~\cite{debug-state}, which defines an action to send a notification message via mobile phone. This action calls the service \codefont{notify.mobile\_app\_huawei\_p20}, with a dynamically generated message to incorporate the reading of a temperature sensor \codefont{sensor.temperatur\_gefrierschrank}. However, as users wrongly accessed the sensor value via \codefont{sensor.temperatur\_gefrierschrank}, the automation does not run and an error message is produced: \emph{``Error rendering data template: UndefinedError: 'sensor ist undefined''}. 

\emph{D4. Wrong format} means that a YAML file is malformed, by violating the formatting rules defined by either YAML or HA. Typical violations include wrong line indentation and wrong quote usage. 

\emph{D5. Wrong predicates} means automation fails because the predicates used in automation have flaws: the defined predicates or predicate combinations do not enable action execution as expected. 
 For instance, Listing~\ref{lst:wrong-algorithm} shows a buggy condition that combines two predicates \codefont{before:sunrise} and \codefont{after:sunset}. The former predicate corresponds to the time period between midnight and sunrise, while the latter corresponds to the period between sunset and midnight. These two predicates should never get used together to define a period ``\emph{after the sunset of Day N and before the sunrise of Day (N+1)}", because the two predicates are always interpreted as two separate periods: ``\emph{after the sunset of Day N and before the sunrise of Day N}'';  no time point satisfies both predicates at the same time.  
% template to define two distinct services for a light template: \codefont{turn\_on} and \codefont{turn\_off}.  Although the services' names imply two opposite operations, their implementations are identical (see lines 13--15 and lines 17--19).
%in the same automation, a trigger fires when an entity's state is changed from ``off'' to ``on'', while the condition requires that entity's state to be ``off'' for one minute. There is no way that the condition can be satisfied after the trigger fires, so the automation never works. 

\begin{listing}
  \caption{A buggy snippet of ``wrong predicates'' (D5)~\cite{wrong-predicates}}  %\vspace{-.5em}
  \label{lst:wrong-algorithm}%\vspace{-.5em}
  \begin{minted}[
    frame=single,
    fontsize=\scriptsize,
    linenos,
    numbersep=2pt,
    breaklines]{yaml} 
# Bug: Wrong combination of predicates    
condition:
  - condition: sun
    before: sunrise
    after: sunset
\end{minted}
\vspace{-.5em}
\end{listing}

\emph{D6. Wrong condition type} is similar to D2. There are multiple types of conditions. However, using some condition type (e.g., device condition) can limit automation expressiveness (e.g., no template allowed), making some logic infeasible.

\emph{D7. Missing predicates} means one or more predicates are not checked before an automation executes certain action. 

\emph{D8. Wrong usage of \codefont{wait\_for\_trigger}}: Action \codefont{wait\_for\_trigger} allows automation to wait for a trigger being fired before doing anything. Some users messed it up with (1) another specialized action \codefont{wait\_template}, which waits for a template to evaluate to true before running automation, or (2) the trigger segment.
%(see Section~\ref{sec:rq2} for more details of these concepts). 

\emph{D9. Flaky program behavior} means two semantically equivalent implementations of the same logic have divergent outcomes: one automation succeeds and the other fails.
%means an automation presents divergent behaviors even if it is implemented in semantically equivalent ways.

%\emph{D9. Wrong condition combination} means condition predicates are not organized correctly. For instance, the conditions combined with an \codefont{and}-operator conflict with each other.

\emph{D10. Visual editor issue} means a problematic YAML file is generated due to the usage of a visual editor in HA IDE.

\emph{D11. Wrong data specification for service calls} means an action calls service(s) using wrong data, making the entire automation fail. Namely, if we treat service calls as analogous to method calls in Java programming, then wrongly specified service data is analogous to wrongly provided parameter values. For instance, Listing~\ref{lst:wrong-service-data} shows a service call, which intends to make the Google speaker announce an audio message \codefont{``Dryer has started''}. However, lines 7--8 are not needed by the service call. When both key-value pairs are provided, the automation fails and an error message is generated: \emph{extra keys not allowed @data['cache\_dir']}.

\begin{listing}
  \caption{A buggy snippet of D11~\cite{wrong-service-data}}  %\vspace{-.5em}
  \label{lst:wrong-service-data}%\vspace{-.5em}
  \begin{minted}[
    frame=single,
    fontsize=\scriptsize,
    linenos,
    numbersep=2pt,
    breaklines]{yaml} 
# Bug: A service call with wrongly provided data    
action:
  - service: tts.google_say
    data:
      entity_id: media_player.living_room_speaker
      message: Dryer has started
      cache: true
      cache_dir: /tmp/tts
\end{minted}
\vspace{-.5em}
\end{listing}

\emph{D12. Wrong action type} is similar to D2 and D6. It happens when users wrongly select a type to implement some unsupported logic.

\emph{D13. Wrong usage of states in action(s)} means users try to define an action by specifying entity states. However, action can be only defined via interactions with services or events (e.g., service calls).

%\emph{D15. Wrong conditional action structure}: The conditional action structure is analogous to the \codefont{switch}-statement in Java. With this structure, users can define an action to call different services when different condition predicates are satisfied. If the structure is wrongly defined, automation fails to work.
%different services called when different conditions are met.
%Action \codefont{choose} is analogous to the \codefont{switch}-statement in Java. When users wrongly define a malformed \codefont{choose} action, automation fails to work.

\emph{D14. Wrong loop structure} means users wrongly use the loop construct to define a malformed automation.

\emph{D15. Wrong usage of \codefont{wait\_template}} means users have difficulty correctly using this action.
 
\emph{D16. Wrong trigger-condition-action structure} means users wrongly define the trigger-condition-action structure to implement an automation, like defining an action without trigger.
 
\emph{D17. Wrong selector used:} Selectors are used to specify what values are acceptable by automation, or define how the input is shown in GUI. D19 is similar to D2, D6, and D12. Basically, when multiple alternative selectors are available, users have difficulty choosing the right one to specify the intended logic.

\emph{D18. Wrong control of the frequency/duration of service calls} is similar to D2, D6, D12, and D17. There are multiple means to control the frequency or duration of service calls, and these means differ in expressiveness. If users choose a wrong mean, then they are unable to express the intended logic.

\emph{D19. Third-party integration issue} means when an automation depends on a third-party software for device control, users may have difficulty fixing issues due to that software usage. 
%fixing bugs introduced by that software usage.

\textbf{The 52 implementation issues cover 13 distinct concepts. Three concepts are shared among segments (I1--I3)}. 

\emph{I1. To retrieve or calculate data} is similar to D3, because it also reflects users' concerns on appropriate data access or calculation. For instance, lines 8--12 in Listing~\ref{lst:retrieval} show a correct snippet from an implementation thread~\cite{impl-state}, which defines an action to send a notification message via mobile phone. The message is formulated based on the value of a sensor that tracks energy production \codefont{sensor.solaredge\_current\_power}. This snippet was recommended because some users could not retrieve the sensor value. Note that I1 is different from D3, as I1 covers implementation questions while D3 is about debugging questions. 

\emph{I2. To specify data for matching} is similar to D1, as it also reflects users' concerns on data specification for predicate-value matching. For instance, lines 8--15 in Listing~\ref{lst:match} show a snippet from an implementation thread~\cite{not-to}. The snippet defines a state trigger, which monitors four sensors (\codefont{sensor.pcX\_printjob}), and starts automation when any sensor (1) changes its state and (2) the to-state is not \codefont{'unknown'}. This snippet was suggested because some users could not specify an unwanted state for the trigger.
%Note that I2 is different from D1, as it characterizes users' request to implement certain features and D1 describes recurring bugs found in users' snippets. 

\emph{I3. To define different service calls} concerns about the correct way of calling distinct services based on the fired triggers, satisfied conditions, or collected data.

\emph{I4. To define varying trigger(s)} means to define one or more changeable triggers (i.e., triggers with variables or templates), whose firing rules vary with the surrounding environment, time, or inputs.

\emph{I5. To define a group-based trigger(s)} means to define triggers based on the values of a group of sensors.

\emph{I6. To define a selection-based trigger} means to define a trigger, which fires based on which GUI button users click. %clicked by users on GUI. 

\emph{I7. To control frequency/duration of service calls} is similar to D18, because it also concerns the correct way of controlling the frequency or duration of service calls. 
 %some users seek for help in choosing and implementing the correct logic to fulfill such automation need.

% for such implementation need in their automation scenario.
%to choose the correct way of controlling the frequency/duration of service calls.
% is similar to D20, as it is also about the correct implementation to control frequency/duration of service calls.

%\emph{I8. To define a conditional action} is similar to D15, as users are also concerned about defining an action, which calls distinct services when different conditional predicates are satisfied.
%means to call different services when different conditions are met.

\emph{I8. To define varying \codefont{wait\_for\_trigger}(s)} is similar to D8, as it also focuses on the correct usage of \codefont{wait\_for\_trigger}.

\emph{I9. To stop a repetitive service call} means when users call a service in a loop structure, they need help in defining the loop so that the service is called repetitively, and the loop condition is properly evaluated before automation terminates.
%ways to specify the frequency or duration or service calls. If users choose a wrong

\emph{I10. To group entities} means to group a set of entities (e.g., speakers) so that they work unanimously just like one entity.

\emph{I11. To call a service based on complex combinations of triggers and conditions} means users have complex predicates/tests to meet, before a service is called. Thus, they specify their automation needs and all predicates, asking for a well-structured automation that properly puts predicates in triggers and conditions.

Both \emph{I12} and \emph{I13} are self-explanatory, so we skip the explanation for succinctness. 

\textbf{The nine optimization issues cover three unique concepts}: to avoid unnecessary trigger firings, to efficiently define changeable triggers, and to reduce duplicated code in automation. One of the concepts is shared among segments (O1).

We use \textbf{segment-agnostic concepts} to refer to the concepts shared among distinct segments, and use \textbf{segment-specific concepts} to refer to the remaining. 
Fig.~\ref{fig:taxonomy} shows one, three, and eight segment-agnostic concepts separately found in optimization, implementation, and debugging issues.
They imply that users ask similar or related questions even though they work on distinct segments.
%These seven concepts are component-agnostic; users make such mistakes even though they work on distinct components. 
%In addition to component-agnostic concepts, issues related to trigger, condition, action, and automation separately cover one, three, six, and four component-specific concepts. 
Additionally, across all categories, 
issues related to trigger cover in total five segment-specific concepts (i.e., D9, I4--I6, O2). Meanwhile, issues related to condition, action, and automation separately cover one, nine, and eight segment-specific concepts. These observations imply that the issues related to trigger and condition are less diverse than those related to action and automation; action and automation are harder to develop and maintain.

\vspace{.2em}
\noindent\begin{tabular}{|p{8.5cm}|}
	\hline
	\textbf{Finding 3:} \emph{One, three, and eight segment-agnostic concepts are separately identified in optimization, implementation, and debugging issues.
This implies the wide existence of common challenges.}\\ \hline
\end{tabular}
\vspace{.2em}

Between the two categories---debugging and implementation,  we observed similarity among identified concepts: (i) D1 vs.~I2, (ii) D3 vs.~I1, (iii) D8 vs.~I8, (iv) D18 vs.~I7, (v) D11 vs.~I3. 
For instance, both D1 and I2 focus on ``data specification for matching''; however, the former describes a root cause of recurring bugs and the latter describes a frequent implementation request. 
Between implementation and optimization, we also observed a commonality: to define varying triggers (I4 vs.~O2). All such similarity or commonality implies that some concepts are challenging and popular; they confuse people no matter whether people implement, debug, or optimize automations.

To facilitate understanding, we use representative examples to show two common challenges. 
Listing~\ref{lst:match} has a buggy snippet from a debugging thread of D1, and a correct snippet from an implementation thread of I2. 
%Each snippet defines a state trigger, which starts an automation when an entity changes from a (specified) state to another. 
%Options like \codefont{from:}, \codefont{to:}, and \codefont{not\_to:} are used to specify states for value matching. 
%For instance, the first trigger tries to monitor a variable \codefont{input\_boolean.ecobee\_fan\_on\_off}, and start automation when the state is changed from \codefont{'off'} to \codefont{'on'}. The other trigger monitors four sensors (\codefont{sensor.pcX\_printjob}), and starts automation when any sensor (1) changes its state and (2) the to-state is not \codefont{'unknown'}. 
%The challenge shown by the buggy fragment is that 
In the first snippet, users committed mistakes by misspelling states (e.g., \codefont{'On'} instead of \codefont{'on'}), and the automation never started. %Unfortunately, users could not fix those errors independently.
The second snippet was suggested because some users could not specify an unwanted state. Both snippets evidence that users have difficulty in specifying data for matching.
%, and they do not have good tool support to help with data specification.
% users have difficulty specifying states for matching. When users misspelled states (e.g., \codefont{'On'} instead of \codefont{'on'}), the automation never starts and users could not detect or fix those errors.
%To illustrate another common challenge, 
Listing~\ref{lst:retrieval} shows another buggy snippet and another correct snippet. In the former one, users made mistakes when accessing a sensor value; the latter one 
%the second snippet 
was recommended because some users asked for help in getting a sensor value.
Both snippets imply that users have difficulty in retrieving or calculating data.

\begin{table*}
\caption{The representative resolution strategies (i.e., each strategy occurs at least four times in our dataset)}\label{tab:resolution}\vspace{-.5em}
\scriptsize
\begin{tabular}{R{.4cm}| p{2.4cm}| p{6.cm}|R{1.cm}| p{6.2cm}}
\toprule
\textbf{Idx}&\textbf{Focus} &\textbf{The Relevant Technical Concept(s)} &\textbf{\# of Issues Resolved}&\textbf{Strategy Content}\\ \toprule
R1 & Quote usage& Wrong format & 10 & Add or remove quotes to properly use String literals, variables, and templates\\ \hline
R2 & Indentation usage& Wrong format & 6 & Add or remove white spaces to properly align statements \\ \hline
R3& Picking a trigger type &Wrong trigger type& 4&Replace the device trigger with a (numeric) state trigger \\ \hline
R4 & Reading an entity's state& Wrong data retrieval or calculation, to retrieve or calculate data & 9& Use template {\tt \{\{states('ENTITY')\}\}} to read the state of ENTITY\\ \hline
R5&Specifying states for matching & Wrong data specification for matching, to specify data for matching & 8&Look up all possible states of the entity in Developer Tools -> States, and use one of the states with case sensitivity.\\ \hline

R6& Different service calls based on fired triggers& To reduce duplicated code, to avoid unnecessary triggering, wrong data specification for matching & 5&Define multiple triggers, specify a unique ID for each trigger, and call services differently based on the fired trigger.\\ \hline

R7&Different service calls based on entity values& To reduce duplicated code, to define different service calls, wrong data specification for matching, wrong loop structure&7 &Define a dictionary to map entity values to distinct service calls or service data, and call services based on the retrieved entity values.\\ \hline

R8& Handling a group or list of same-typed entities & To reduce duplicated code, to define different service calls, to define varying trigger(s), wrong data retrieval or calculation, wrong algorithm design & 5 & Use {\tt expand()} to enumerate a group or list of entities, and define filters to pick elements that satisfy certain requirements. \\ \bottomrule
\end{tabular}
\vspace{-2.2em}
\end{table*}
%could not properly access sensor values. The second snippet was recommended 
%The challenge here is that users have difficulty in accessing sensor values. In particular, the buggy fragment implies that when users erroneously accessing sensor values, there is insufficient tool support to help them fix the error. 

\vspace{.2em}
\noindent\begin{tabular}{|p{8.5cm}|}
	\hline
	\textbf{Finding 4:} \emph{Five concepts commonly exist in implementing, debugging, and optimization issues; these concepts are challenging no matter whether users implement, debug, or optimize automations. %These concepts are challenging perhaps due to the lack of documentation and tool support.
	}\\ \hline
\end{tabular}
%\vspace{.2em}

\subsection{RQ2: Frequent Resolution Strategies}\label{sec:rq2}

We realized that the \totalInspected examined issues were resolved in diverse ways; no dominant strategy was applied to resolve the majority.
However, we managed to identify some strategies repetitively applied to resolve multiple issues  ($\ge$ 4). 
Table~\ref{tab:resolution} lists each strategy in terms of its focus, the relevant technical concept(s), the number of issues it addressed, and the strategy content.
We ranked the eight strategies in ascending order of the number of relevant concepts.  
 This section explains all strategies in detail.

%# Bug 1    
%service: timer.start
%target:
%  entity_id: {{ states('input_select.select_cooking_timer') }}
%# Fix 1: to quote the template, as shown below
%  entity_id: ''{{ states('input_select.select_cooking_timer') }}''

%# Bug 3
%service: input_text.set_value
%data:
%  entity_id: input_text.telegram_disparador_alarma
%  value: >
%    {% if is_state('trigger.platform', 'time') %}
%# Fix 3: to remove the quote around trigger.platform, as below
%    {% if is_state(trigger.platform, 'time') %}

\textbf{R1} corrects quote usage. It was frequently applied when askers posted YAML files with wrong formats.
For instance, Listing~\ref{lst:quote} shows a representative bug and fix. The buggy snippet tries to set a counter \codefont{counter.configure} using the value of another counter \codefont{counter.aux\_ac\_pieza}, but the quote usage violates a domain-specific constraint in HA: when there are quotes inside and outside a template, differentiate the outer and inner ones by using different quote types. 
The buggy code nests single quotes. Thus, the fix is to replace the inner pair of single quotes with double quotes.
% the domain-specific constraints in HA. 
%Namely, HA has domain-specific constraints on quote usage, and the violation of any constraint can fail automation configurations. 
%Exemplar HA constraints on quote usage include: 
\begin{comment}
\begin{itemize}
\item[(c1)] Templates should be wrapped with single or double quotes. 
\item[(c2)] Inside a template, literals (e.g., entity name \codefont{counter.aux\_ac\_pieza} in Listing~\ref{lst:quote}) should be quoted. 
\item[(c3)] When there are quotes outside and inside a template, differentiate the outer and inner ones by using different quote types. For instance, wrap a template with double quotes and wrap all literals inside it with single quotes, or vice versa.
\end{itemize}
\end{comment}

\begin{listing}[b]
\vspace{-1.5em}
  \caption{A bug and fix related to the quote usage (R1)~\cite{quote}}\label{lst:quote}
    \vspace{-.5em}
  \begin{minted}[
    frame=single,
    fontsize=\scriptsize,
    linenos,
    numbersep=2pt,
    breaklines]{yaml}        
service: counter.configure
data_template:
  value: '{{states('counter.aux_ac_pieza')}}'
# Bug: A pair of single quotes enclose another pair of single quotes  
# Fix: Replace the inner quotes with double quotes, as below
  value: '{{states(''counter.aux_ac_pieza'')}}'
  \end{minted}
  \vspace{-.5em}
\end{listing}
\begin{listing}[b]
\vspace{-1.5em}
\caption{A bug and fix about indentation (R2)~\cite{indent}}
\label{lst:indent}
\vspace{-.5em}
\begin{minted}[
    frame=single,
    fontsize=\scriptsize,
    linenos,
    numbersep=2pt,
    breaklines]{yaml} 
action:
  - repeat:
      while:
        - condition: state
          entity_id: binary_sensor.group_door_sensor_at_night
          state: 'on'
        - condition: time
          before: '05:00:00'
          after: '23:30:00'
sequence:
  - service: notify.mobile_app_iphone
# Bug: The while-loop does not have "sequence" aligned with "while"      
# Fix: To indent the "sequence" block
      sequence:
        - service: notify.mobile_app_iphone
  \end{minted}
  \vspace{-.5em}
\end{listing}

\textbf{R2} corrects indentation. As with R1, this was also adopted to fix wrong formats. 
% It was also often applied when askers posted YAML files with wrong formats.
The basics of YAML syntax are block collections and mappings containing key-value pairs. %Each item in a collection starts with a ``\codefont{-}'', while mappings have the format "\codefont{key:value}". 
Indentation is important for specifying relations among collections, mappings, and their items.
% Indented lines are nested inside lines that are one level higher.
 For instance, the buggy version in Listing~\ref{lst:indent}
 tries to repetitively (1) 
 check whether the door is open at night (between 23:30pm and 5am next day), and (2) send a notification if so. Such a logic can be realized with a \codefont{while}-loop, which encloses two blocks---a \codefont{while} block and a \codefont{sequence} block---with keyword \codefont{repeat}. However, as the buggy version did not indent the \codefont{sequence} block properly (see lines 10--11), the automation fails. 
 %The fix should change the indentation of \codefont{sequence}-block, and align it with the \codefont{while}-block. 
 % if users did not indent the \codefont{sequence} block correctly 

\textbf{R3} replaces device triggers with triggers of numeric\_state or state, because device triggers are too limited to express triggering events. For instance, the buggy version in Listing~\ref{lst:trigger} defines a device trigger, to fire when sensor \codefont{binary\_sensor.v4b01\_motion} detects no motion for the number of minutes users specified via \codefont{input\_number.timeout\_offices}, a numeric input box in GUI.
However, device triggers do not support templating (see line 11), so the buggy version fails. 
The fix is to replace that trigger with a state trigger (lines 14--20).
%a state trigger to replace the device trigger.
% and can express the same logic in a different way.

\textbf{R4} is about state access. To correctly read the state of an entity (see Listing~\ref{lst:retrieval}), people are suggested to call function \codefont{states(...)} with that entity's ID, such as \codefont{states('sensor.solaredge\_current\_power')}.

\textbf{R5} fixes misspelled states, by suggesting users to look up valid states of a given entity in Developers Tools of HA IDE and properly specify states. For example, \codefont{input\_boolean.ecobee\_fan\_on\_off} in Listing~\ref{lst:match} is a boolean variable representing an input box in GUI, whose state value is \codefont{'on'} or \codefont{'off'}. Users must specify states accordingly.
%As HA is case sensitive, the misspelled states (see lines 5--6) must be fixed.
%result in a broken trigger and the fix corrects state specification.
%When the snippet misspells states as \codefont{'On'} and \codefont{'Off'}, 
%the buggy snippet in Listing~\ref{lst:match} defines a state trigger with misspelled states \codefont{'On'} and \codefont{'Off'}. The entity \codefont{input\_boolean.ecobee\_fan\_on\_off} is a boolean input box in GUI, whose state is \codefont{'on'} or \codefont{'off'}.
%which should fire when an \codefont{input\_boolean} helper is updated from the \codefont{'off'} state to \codefont{'on'}. Because HA is case sensitive, the misspelled states (see lines 5--6) result in a broken trigger.

\textbf{R6} defines alternative triggers in one automation rule, and calls services differently depending on the fired trigger. 
It establishes correspondence between triggers and service calls by (1) assigning distinct IDs to triggers, and (2) referring to those IDs in service calls. For instance, suppose that an HA user wants to turn on a wifi bulb at 3pm and off at 5pm every day; the bulb should be also turned on whenever it gets online during 3pm-5pm after being offline, because the bulb sometimes goes offline at 2:59pm due to connectivity issues. 
To define both turning-on and turning-off behaviors of the bulb in the same automation rule,   lines 2--12 in Listing~\ref{lst:trigger-ids} define three alternative triggers, and assign two IDs to those triggers separately: \codefont{'on'} and \codefont{'off'}; line 15 calls the corresponding service (i.e., \codefont{light.turn\_on} or \codefont{light.turn\_off}) by composing a service name using the \codefont{trigger.id} value. 

%an HA user wants to turn on a bulb at 3pm and off at 5pm every day. Because the bulb sometimes goes offline at 2:59pm and goes online at 3:02 for certain reasons (e.g., connectivity), the HA user also wants the bulb \codefont{light.mybulb} 
%the automation tries to turn on a bulb at 3pm and off at 5pm every day. Because the bulb sometimes goes offline at 2:59pm

\begin{listing}[t]
%\vspace{-.5em}
  \caption{Different service calls based on fired triggers (R6)~\cite{exceptions}}\label{lst:trigger-ids}
   %\vspace{-1.5em}
  \begin{minted}[
    frame=single,
    fontsize=\scriptsize,
    linenos,
    numbersep=2pt,
    breaklines]{yaml} 
- alias: TurnOnOffBulb
  trigger:
    - id: 'on'
      platform: time
      at: '15:00:00'
    - id: 'off'
      platform: time
      at: '17:00:00'
    - id: 'on'
      platform: state
      entity_id: light.mybulb
      from: 'unavailable'
  condition: "{{ today_at('15:00:00') <= now() < today_at('17:00:02') }}"
  action:
    - service: 'light.turn_{{ trigger.id }}'
      target:
        entity_id: light.mybulb
\end{minted}
\vspace{-.5em}
\end{listing}

%\begin{comment}
\begin{listing}[b]
\vspace{-.5em}
  \caption{Different service calls based on entity values (R7)~\cite{automation-reference}}\label{lst:entity-values}
   \vspace{-1.5em}
  \begin{minted}[
    frame=single,
    fontsize=\scriptsize,
    linenos,
    numbersep=2pt,
    breaklines]{yaml} 
- id: '1631649587197'
  alias: Alert - Bin collection tomorrow
  description: ''
  variables:
    waste: >
      {% set t = (now() + timedelta(days=1)).date() %}
      {% set x =
 { strptime(states('sensor.rubbish_bin_collection'),'%a %d %b %Y').date(): 'Rubbish', strptime(states('sensor.recycling_bin_collection'),'%a %d %b %Y').date(): 'Recycling', strptime(states('sensor.garden_bin_collection'),'%a %d %b %Y').date(): 'Garden' } %}
      {{ x[t] if t in x.keys() else 'nothing' }}
  trigger:
  - platform: time
    at: '20:00:00'
  condition: "{{ waste != 'nothing' }}"
  action:
  - service: notify.admin_devices
    data:
      message: "{{ waste }} bin collection tomorrow!"
  mode: single
\end{minted}
  \vspace{-.5em}
\end{listing}
%\end{comment}

\textbf{R7} reads values of multiple entities (e.g., sensors), and calls services differently depending on those values. It typically defines a dictionary to map entity values with service calls or data. Given an entity value, it looks up the dictionary to take an action.
For instance, Listing~\ref{lst:entity-values} uses three self-defined sensors to pull bin collection dates (rubbish, recycling, garden) from a website, and sends notification messages the night before a collection is due. In the automation, the Jinja variable \codefont{x} is defined as a dictionary (line 7), to map pulled dates of bin collection to bin types. The Jinja variable \codefont{t} holds the date of next day (line 6). Variable \codefont{waste} is initialized as \codefont{'nothing'} if the next day's date \codefont{t} does not match any bin collection date, or as a bin type if \codefont{t} finds a match (line 9). Finally, the service is called with the value set to \codefont{waste} (line 17).

\textbf{R8} defines pipelines to uniformly handle a group or list of same-typed entities. Each pipeline typically starts with the Jinja function call \codefont{expand()} to enumerate all elements in a group/list, and  then invokes Jinja built-in filters (e.g., \codefont{map()}) to refine or process elements. For instance, suppose that an HA user wants to set all Google speakers at home to the same volume, by referring to the most recent volume setting among all speakers.
 Listing~\ref{lst:group-entities} satisfies the automation need through three steps. First, to identify all turned-on speakers, it enumerates speakers  (lines 4--7), rejects turned-off entities (line 7), gathers values of attribute \codefont{'entity\_id'} (line 8), and lists those values. Second, to acquire the most recent volume setting, it enumerates all volume sensors (lines 10--13), sorts them in descending order of their latest change timestamps (line 14), gathers values of attribute \codefont{'state'} (i.e., timestamps), gets the first one in that list (i.e., the mosts recent one), and converts the value to a floating-point number. Third, the automation sets all turned-on speakers to the same volume. 

\begin{listing}[b]
\vspace{-.5em}
  \caption{Handling a group or list of same-typed entities (R8)~\cite{automation-action}}\label{lst:group-entities}
   \vspace{-1.5em}
  \begin{minted}[
    frame=single,
    fontsize=\scriptsize,
    linenos,
    numbersep=2pt,
    breaklines]{yaml} 
action:
  - variables:
      speakers_on: > 
        {{ expand("media_player.google_home_bedroom",
        "media_player.google_nestmini_office", 
        "media_player.google_nesthub_living_room", 
        "media_player.google_nestmini_kitchen")| rejectattr('state', 'eq', 'off')
        | map(attribute='entity_id') | list }}
      recent_volume: >
        {{ expand("sensor.volume_google_speaker_bedroom",
        "sensor.volume_google_speaker_office",
        "sensor.volume_google_speaker_living_room",
        "sensor.volume_google_speaker_kitchen" ) 
        | sort(attribute='last_changed', reverse=true)
        | map(attribute='state') | first | float(0) }}
  - service: media_player.volume_set
    data:
      volume_level: "{{ recent_volume }}"
    target:
      entity_id: "{{ speakers_on }}"
\end{minted}
  \vspace{-.5em}
\end{listing}

Notice that
R1--R3 address debugging issues; R4--R5 resolve debugging and implementation issues; R6 fixes optimization and debugging issues; R7--R8 handle issues of all three categories. These strategies provide two insights for future tool support of smart home development. 
First, users repetitively commit certain mistakes, 
and such mistakes present clear bug patterns; it is promising to create new tools to detect and fix repetitive bugs. 
%that detect and fix repetitive bugs, in order to help improve software quality and programmer productivity. 
Second, some users cannot (1) define diverse actions depending on the fired triggers or entity values, or (2) define uniform processing for elements in a group/list, although such automation needs are not uncommon. Therefore, it will be helpful to create tools, which generate automation implementations to at least partially satisfy those needs.

\vspace{.2em}
\noindent\begin{tabular}{|p{8.5cm}|}
	\hline
	\textbf{Finding 5:} \emph{The eight resolution strategies imply that domain experts addressed recurring issues by following certain principles; it is promising to automate such principles.
	%Six concepts commonly exist in implementing, debugging, and optimization issues. These concepts are challenging perhaps due to the lack of documentation and tool support.
	}\\ \hline
\end{tabular}

%First, developers/users may repetitively occur certain mistakes (e.g., wrong quote or indentation usage), 

\vspace{-.5em}
\subsection{RQ3: Current Debugging Tool Support}\label{sec:rq3}

To study the existing tools for debugging YAML files or YAML-based automation configuration, we searched online with keywords ``YAML validator'' and found five publicly available tools: YAML Validator by Code Beautify~\cite{codeBeautify_yaml}, YAML Lint~\cite{yaml_lint}, YAML Checker~\cite{yaml_checker}, YAML Validator by JSON formatter~\cite{json_formatter_yaml}, and ONLINEYAMLTOOLS~\cite{online_yaml_tools}. Additionally, as HA provides an IDE to help automation development, we also included its checker into the tool list for evaluation. As shown in Table~\ref{tab:tools}, for simplicity, we assigned a unique ID to each tool, and will consistently use T1--T6 to refer to these tools. In our experiments, \textbf{because we found no tool to suggest any bug fix, this section focuses on our results about tools' capabilities of bug detection.}

\subsubsection{Metrics for Automatic Bug Detection} We defined three metrics to evaluate the bug detection capabilities of tools:

\emph{\textbf{Recall (R)}} measures among all known bugs, how many of them are reported by a tool:
\begin{equation}
R = \frac{\text{\# of known bugs reported}}{\text{Total \# of known bugs}}
\end{equation}
To compute the recall of a tool, we first intersected the tool-reported bugs with known \totalDebug bugs. Then we computed the count ratio between this intersection and the known bug set.

\textbf{\emph{Precision (P)}} measures among all bugs reported by a tool, how many of them are real bugs: 
\begin{equation}
P = \frac{\text{\# of true bugs reported}}{\text{Total \# of bugs reported}}
\end{equation}
The 129-bug set does not include all bugs existent in given YAML files/snippets, as developers sometimes omitted discussion on trivial bugs. To compute a tool's precision, we manually inspected all bug reports. If a  report $B_r$ clearly describes a known bug,
 we consider $B_r$ a true positive. Otherwise, if $B_r$ is confusing, we applied the tool which originally output $B_r$ also to the developer-fixed version. If (1) that tool reported nothing for the fixed version or (2) our domain knowledge confirms the bug reported by $B_r$, we consider $B_r$ a true positive; otherwise, $B_r$ is a false positive.

\textbf{\emph{F-score (F)}} combines \emph{P} with \emph{R}, to measure the overall accuracy of bug detection as below:
\begin{equation}
F = \frac{2 \times P \times R}{P + R}
\end{equation}
\emph{F} is the harmonic mean of \emph{P} and \emph{R}. All three metrics have their values vary within [0, 1]. The higher, the better.

\begin{table}
\scriptsize
\centering
\caption{The bug detection capabilities of current tools}%Existing tool support that help debug automation configurations}
\label{tab:tools}
\vspace{-1.em}
\begin{tabular}{l| p{2.cm}| rr| rrr}
\toprule
\multicolumn{1}{c|}{\multirow{2}{*}{\textbf{ID}}} & \multicolumn{1}{c|}{\multirow{2}{*}{\textbf{Tool Name}}} & \multicolumn{2}{c|}{\textbf{\# of Bugs Reported}} &\multirow{2}{*}{\textbf{P}} &\multirow{2}{*}{\textbf{R}} &\multirow{2}{*}{\textbf{F}}\\
\multicolumn{1}{c|}{} & \multicolumn{1}{c|}{} & Correct & Incorrect\\
\toprule
T1 & YAML Validator by Code Beautify \cite{codeBeautify_yaml} & 15 & 3 &83\% & 9\% &15\%\\ \hline
T2 & YAML Lint \cite{yaml_lint} & 14 & 3 &82\% &7\% &13\%\\\hline
T3 & YAML Checker \cite{yaml_checker} & 15 & 5 &75\% &8\% &14\% \\\hline
T4 & YAML Validator by JSON formatter~\cite{json_formatter_yaml} & 15 & 4&79\%&8\%&14\%\\\hline
T5 & ONLINEYAML\- TOOLS~\cite{online_yaml_tools} & 15&4&79\%&8\%&14\%\\ \hline
T6 & HA IDE Checker \cite{ha} & 15&1&94\%&8\%&14\%\\
\bottomrule
\end{tabular}
\vspace{-1.5em}
\end{table}

\subsubsection{Tool Effectiveness}
As shown in Table~\ref{tab:tools}, the tools behaved similarly to each other. All tools achieved high precision (75\%--94\%), low recall (7\%--9\%), and low F scores (13\%--15\%).
They reported bugs in 14 common YAML files/snippets, and detected no bug in 106 common files/snippets. 
%They commonly reported bugs in 13 YAML files/snippets, and overlooked the bugs in 
High precision means that the six tools report bugs with low false positives. Namely, if a tool reports a bug for a given YAML file/snippet, the file or snippet is very likely to be buggy. Meanwhile, low recall means that these tools poorly reveal known bugs. Among the \totalDebug bugs we distilled from HA discussion, only 9--11 bugs were covered by the reports generated by each tool; the combination of all tools' outputs only revealed 14 known bugs. 
%by tool-generated reports; the remaining 119--121 ones were undetected and tools wrongly treated those files/snippets as valid ones. Even if we combined all tools' outputs, the combination only revealed 14 of the known bugs. 
Due to such low recall rates, the measured F values are also low.

Among the tools, T6 got the highest precision; T1 got the highest recall and F-score; T4 and T5 acquired identical values for all metrics.  
Between T4 and T5, we found a significant overlap in tool-generated bug reports; both tools described all errors in almost identical ways. This interesting observation implies that T4 and T5 may share the same core implementation. Additionally, we once hypothesized T6 to outperform other tools in all metrics, because T6 was specially designed to reveal HA-related issues while all other tools are general YAML file checkers. Surprisingly, we found T6 to work similarly with other tools, and only outperformed the other tools in precision.
%It only slightly outperformed T4 and T5, but worked slightly worse than another three tools.

\vspace{.2em}
\noindent\begin{tabular}{|p{8.5cm}|}
	\hline
	\textbf{Finding 6:} \emph{
	For bug detection, existing tools achieved high precision, low recall, and low F scores.
	}\\ \hline
\end{tabular}

\begin{comment}
\begin{listing}[b]
  \vspace{-1.em}
  \caption{A buggy snippet and its fix}\label{lst:error}
    \vspace{-1.em}
 \begin{minted}[
    frame=single,
    fontsize=\footnotesize,
    linenos,
    breaklines]{yaml} 
service: counter.configure
data_template:
  value: '{{states('counter.aux_ac_pieza')}}' # Bug
# Fix: Replace the inner single quotes with double quotes
  value: '{{states(''counter.aux_ac_pieza'')}}'
  \end{minted}
  \vspace{-1.5em}
\end{listing}
\end{comment}

\subsubsection{Characterization of Tool-Generated Bug Reports} Among the bug reports output by different tools, we observed two phenomena. 
First, \emph{the bug reports are purely about wrong formats.} These issues may involve wrong quote usage, bad indentation, and wrong tag (i.e., token) usage. However, as mentioned in Section~\ref{sec:rq1}, the \totalDebug known bugs cover \totalConcepts distinct concepts; ``wrong format'' is just one of them, and its bugs seem easier to detect because formatting rarely requires for advanced program analysis. %is often irrelevant to program semantics. 
Unfortunately, existing tools cannot detect other types of bugs. 
Even within the 20 known bugs of ``Wrong format'', we still observed 11 cases missed by all tools. For three cases, T3--T5 failed to interpret the legal notation ``!'' and incorrectly reported it as wrong tag usage.

\begin{table}[t]
\caption{The confusing tool-generated reports for Listing~\ref{lst:quote}}\label{tab:error}
\vspace{-1.em}
\footnotesize
\begin{tabular}{l| p{7.5cm}}
\toprule
\textbf{ID} &\textbf{Error Message}\\ \toprule
T1 & Error: Unexpected characters near "counter.aux\_ac\_pieza')\}\}'". Line : 3 value: '\{\{states('counter.aux\_ac\_pieza')\}\}'\\ \hline
T2 & Unexpected scalar at node end at line 3, column 21\\ \hline
T3 & bad indentation of a mapping entry (3:21)\\ \hline
T4 & Error: can not read an implicit mapping pair; a colon is missed at line 3, column 46\\ \hline
T5 & YAMLException: can not read an implicit mapping pair; a colon is missed at line 3, column 46\\ \hline
T6 & bad indentation of a mapping entry (3:21)\\ 
\bottomrule 
\end{tabular}
\vspace{-1.5em}
\end{table}

\vspace{.2em}
\noindent\begin{tabular}{|p{8.5cm}|}
	\hline
	\textbf{Finding 7:} \emph{Existing tools only revealed a relatively simple type of bugs---wrong format; even for the detection of such bugs, current tools suffer from significant false positive and false negative issues.
	}\\ \hline
\end{tabular}
\vspace{.3em}

Second, \emph{the error messages in many bug reports are very confusing.} They may incorrectly pinpoint the bug location, wrongly describe an error, or provide meaningless hints that mislead users. For instance, Listing~\ref{lst:quote} shows a buggy snippet to misuse quotes at line 3. Essentially, single quotes 
should not be nested. 
%; the inner pair of single quotes should be replaced with double quotes.
However, T1 and T2 output confusing messages to complain about unexpected characters/scalar at line 3 (see Table~\ref{tab:error}), without clarifying which character/scalar is unexpected.
T3 and T6 reported bad indentation, but no indentation issue exists at all.
%although no indentation issue exists at all. 
T4 and T5 mentioned a missing colon, which does not reflect the wrongly used quotes, either. 
Among the 109 bug reports we examined in total, 45 reports contain confusing error messages. T1 produced the biggest number of confusing reports---12, while T3--T6 output the fewest---6.

\vspace{.2em}
\noindent\begin{tabular}{|p{8.5cm}|}
	\hline
	\textbf{Finding 8:} \emph{In our experiment, 41\% (45/109) of the examined bug reports are confusing, which evidences the need of improving the error-reporting mechanism.
	}\\ \hline
\end{tabular}
%\vspace{.3em}

\vspace{-.5em}
\section{Threats to Validity}

\emph{Threats to External Validity.} All our observations are based on the experimental dataset. 
Although the study analyzes \totalInspected threads in depth, the sample size, while substantial, may not fully represent the diverse range of experiences and challenges encountered by all users.
%Our observations may not generalize well to the other automation-related discussions on HAC that are not covered by this dataset. 
In the future, we will add more data to our dataset, and conduct further research as well as analysis to draw more generalized conclusions about automation configuration issues in smart homes. 
%Our observations may not generalize well to the other automation-related discussions on HAC that are not covered by this dataset. In the future, we will add more data to our dataset, to make our study more representative.%discussion threads into the dataset, so that our findings are more representative.

\emph{Threats to Internal Validity.} Our manual analysis for the collected data and tool outputs is subject to human bias, and is limited to our domain knowledge. To mitigate the problem, we had two authors independently examine the data in multiple iterations. When they disagreed upon certain data labels or classification criteria, they had discussion until coming to an agreement and enforced the same labeling/classification mechanism in the next iteration.

\vspace{-.5em}
\section{Lessons Learned}

Below are actionable items we learned from this study.

\vspace{.5em}
\emph{\textbf{For Tool Developers and SE Researchers:}} 
%\subsection{Lessons for Tool Builders and SE Researchers}
%HA is widely used; its users and active installations have grown rapidly over the years~\cite{haa}. However, 
Our study shows that HA users do not have sufficient domain knowledge or good tool support for automation development. Tool developers and SE researchers can build tools to (1) better detect or fix bugs in user-defined automations or (2) generate automation implementation from scratch. 
%In this way, they will improve the productivity of end-user programming, strengthen the quality of smart homes, and profoundly influence the future of IoT programming as well as smart community. 
In particular, we suggest three future directions.

First, syntax checkers and fixers. Users often use quotes and indentation in wrong ways (Sections~\ref{sec:rq1} and~\ref{sec:rq2}), and the current tool support is poor (Section~\ref{sec:rq3}). HA defines a set of grammar rules on top of the basic YAML rules (e.g, \codefont{while}-loop). To enforce these domain-specific rules, future work can create parsers to analyze YAML files, locate HA-specific keywords as well as structures, comprehend automation rules based on the extracted information, and create syntax trees. In the tree-creation procedure, parsers can detect malformed content by reporting any violation of grammar rules; they can further automate or suggest fixes by observing grammar rules and program context.

Second, semantic checkers and fixers. As described by Section~\ref{sec:rq2}, HA defines semantic rules on segment-specific concepts (e.g., device triggers do not support templating), and segment-agnostic ones (e.g., sensor states should be accessed via the function \codefont{state(...)}). To enforce these rules, future work can create analyzers to traverse the parsing trees mentioned above, gather semantic information like the definition or usage of variables/templates/entities, compare the gathered information with predefined bug patterns, and report a bug for each found pattern-match. The analyzers can also automate or suggest fixes based on predefined bug-fixing patterns, or data analysis of correct automations.

Third, generators of automation configurations. As mentioned in Section~\ref{sec:rq2}, some users have similar automation needs (e.g., to call distinct services based on the fired triggers), and the code to satisfy those needs share commonality (e.g., defining and using trigger IDs). Future work can use data-driven approaches like machine learning and large language models (LLMs) to (1) infer the correspondence between automation needs and YAML code, and (2) generate YAML code given automation specification. Such tools can be used in combination with the bug detectors and fixers mentioned above, to iteratively refine or optimize automations. 

\vspace{.5em}
\emph{\textbf{For HA Developers:}} Users are often confused about the correct usage of some program constructs, built-in functions, or templates (Section~\ref{sec:rq1}). The error messages generated by existing HA checker seem not quite helpful (Section~\ref{sec:rq3}). To help users better adopt HA and further broaden the platform's impact, HA developers may need to improve  documentation, and to combine existing concrete YAML examples with a more systematic and comprehensive concept explanation. They may also need to enhance the error-reporting mechanism, to clarify the root causes or even fixes of reported bugs. 

\vspace{.5em}
%\paragraph
\emph{\textbf{For HA Users:}} Carefully read the available HA documentation to follow best practices and avoid the well-known pitfalls. When asking questions on HA, provide all relevant information (e.g., automation specification, unsuccessful trials, error messages, and help request) to benefit most from the community wisdom.

\vspace{-.5em}
\section{Related Work}

The related work includes empirical studies on Internet-of-Thing (IoT) systems, and bug detection in those systems.
% new tools to detect bugs in such systems.

\subsection{Empirical Studies on IoT Systems}
Studies were recently performed to characterize issues or problems in IoT systems~\cite{fernandes2016security,He2019,alrawi2019sok,Brackenbury2019,Zhou2021,makhshari2021iot,wang2022understanding}. 
For instance, Fernandes et al.~\cite{fernandes2016security}, Alrawi et al.~\cite{alrawi2019sok}, and Zhou et al.~\cite{Zhou2021} analyzed the security properties of IoT platforms and systems. 
%Zhou et al.~\cite{Zhou2021} reviewed the 20 security bugs mentioned in IoT literatures and classified them. However, it is unknown on which platforms the bugs exist, and whether the list of bugs or categories are representative.%Specifically, Fernandes et al.~leveraged static code analysis and testing to reveal the design flaws of overprivilege  in SmartApps. Alrawi et al.~reviewed existing literature on IoT security vulnerabilities to characterize attack techniques, mitigations, and stakeholders. 
In contrast, our research focuses on coding issues in IoT automation configuration, covering three categories: (1) implementing new features, (2) debugging, and (3) optimization. 
% functional bugs in IoT automation configurations/implementations, instead of security issues.
He et al.~\cite{He2019} did an online survey with 72 users of smart home systems, to learn their negative user experiences. The participants reported fears of breaking the system by writing code, and struggles of diagnosing or recovering from system failures. Similarly, Makhshari and Mesbah~\cite{makhshari2021iot} did interviews and surveys with IoT developers; they also found testing and debugging as the major challenges. %faced by IoT developers.
 However, neither study examines any widely used smart home platform to characterize the bug patterns or fixing strategies, let alone to provide concrete actionable advices to tool builders. % for those platforms.
Our study is motivated by and complement both studies. %and compliment them. 
%Motivated by the observation that IoT developers had lots of difficulty creating smart home systems, we decided to focus on HA---one of the most popular smart home platforms---and characterized the bug patterns as well as fixing strategies.

Brackenbury et al.~\cite{Brackenbury2019} focused on the trigger-action programming (TAP) model. They systematized the temporal paradigms through which TAP systems could express rules, and classified TAP programming bugs into three categories: control logic, timing, and inaccurate user expectation.
As with Brackenbury et al., we also identified bugs related to these three general categories (e.g., wrong loop structure and wrong data specification for matching). 
%characterized bugs related to the TAP model. 
However, our taxonomy is finer-grained and more comprehensive, as we derived bug patterns from real-world bugs instead of speculation on the TAP model. Our observations reflect the real-world bug distribution among patterns; 
%our observations are not limited to the 10 types of bugs
we also characterized (1) bugs violating syntactic rules (e.g., wrong formats) and more diverse semantic rules (e.g., data access), (2) developers' bug fixes, (3) recurring implementation requests, and (4) frequent optimization needs.

%They also conducted an online survey with 153 participants, and found that the presence of a bug made it harder for participants to 
% Furthermore, they conducted an online survey with 153 participants, who were asked to assess a series of pre-written TAP rules and half of the rules exhibited bugs. For most of the bug classes, the authors found that the presence of a bug made it harder for participants to correctly predict the behavior of the rule.  

%Zhou et al.~\cite{Zhou2021} reviewed the 20 security bugs mentioned in IoT literatures and classified them. However, it is unknown on which platforms the bugs exist, and whether the list of bugs or categories are representative.
Wang et al.~\cite{wang2022understanding} inspected 330 device integration bugs mined from HAC to characterize any root cause, fix, trigger condition, and impact of those bugs. Our study is closely related to the work by Wang et al., as we also examined data mined from HAC. However, our study is irrelevant to device integration; instead, it focuses on %automation configuration, aiming to reveal challenges and opportunities in the development procedure of automation configuration.
the coding issues in automation configuration.%, instead of device integration bugs.

\subsection{Automatic Bug Detection in IoT Systems}
Tools were created to automatically detect bugs in IoT systems~\cite{liang2016understanding,celik2018soteria,trimananda2020understanding,Li2020,Xiao2020,Fu2021,Huang2023}. 
Specifically, VulHunter~\cite{Xiao2020} detects new vulnerabilities by analyzing the known vulnerability patch packs in Industry IoT. SOTERIA~\cite{celik2018soteria} verifies whether IoT apps adhere to the identified safety, security, and functional properties via model checking.
Liang et al.~\cite{liang2016understanding} and Fu et al.~\cite{Fu2021} separately created tools to reveal bugs in IoT operating systems (OSes). 
%In particular, Liang et al.~inspected 23 fixed bugs in 3 IoT OSes, summarized 4 bug patterns, and created a rule-based bug detector accordingly. Fu et al.~created a tool CPscan, to identify bugs introduced by developers when they customized the Linux kernel via code pruning to create IoT kernels. CPscan compares the kernels before and after pruning, locates deleted security operations (DSO), and assesses the security impact. 

Some tools~\cite{trimananda2020understanding,Li2020,Huang2023} detect conflicting interactions between smart home IoT applications, because these conflicts can result in undesired actions like locking a door during a fire. For instance, Trimananda et al.~\cite{trimananda2020understanding} studied 198 official and 69 third-party apps on Samsung SmartThings, and found 3 major categories of app conflicts. Based on their observations, the researchers created a conflict detector that uses model checking to detect up to 96\% of the conflicts.
Li et al.~\cite{Li2020} categorized conflicts in a totally different way. They also invented a graph structure named \emph{IA graph} to represent the controls in each IoT app and event schedules. With that representation, they created an efficient algorithm to leverage first-order logic and SMT solvers to detect conflicts. 

Our study complements all tools mentioned above; it focuses on a different type of IoT bugs---bugs in automation configuration or implementation. Such implementation bugs are less relevant to security, IoT kernels, or inter-app conflicts. However, they are still important as these bugs can prevent end-users from realizing the desired automation rules or achieving high-quality automations.

%Fu et al.~created a tool to detect bugs in IoT kernels, which were customized by developers via removing code from the Linux kernel.
% studied 23 fixed bugs in 3 IoT operating systems (OSes): Contiki, TinyOS, and RIOT OS. Based on the root causes summarized for these 23 bugs, the researchers implemented a rule-based bug detector and successfully found previously unknown but later confirmed bugs in those OSes.

\vspace{-.5em}
\section{Conclusion}

%Smart devices are becoming more prevalent in daily life. Correspondingly, 
As the growing prevalence of smart homes, we believe that they will
%the smart homes built on top of smart devices will
 become crucial to lower utility costs, improve people's life, and protect people's properties. Wrongly configured home automations can waste utilities, jeopardize people's life, and compromise home safety/security. Our study characterizes the challenges and opportunities in HA-based smart homes, enlightening
 %(1) the major root causes of technical issues HA users discussed, (2) repetitively applied resolution strategies for issues, and (3) limitations of existing tools support.
%Our will enlighten 
future research to improve end-user programming and smart-home quality.

%\section*{Acknowledgment} We thank anonymous reviewers for their valuable comments on our earlier version of the paper. This work was supported by NSF-1845446 and NSF-1929701.
% \bibliographystyle{ACM-Reference-Format}
% \bibliography{ying-ase-2021}
\bibliographystyle{IEEEtran}
\bibliography{anik-issta-2024}

\begin{thebibliography}{10}
\providecommand{\url}[1]{#1}
\csname url@samestyle\endcsname
\providecommand{\newblock}{\relax}
\providecommand{\bibinfo}[2]{#2}
\providecommand{\BIBentrySTDinterwordspacing}{\spaceskip=0pt\relax}
\providecommand{\BIBentryALTinterwordstretchfactor}{4}
\providecommand{\BIBentryALTinterwordspacing}{\spaceskip=\fontdimen2\font plus
\BIBentryALTinterwordstretchfactor\fontdimen3\font minus
  \fontdimen4\font\relax}
\providecommand{\BIBforeignlanguage}[2]{{%
\expandafter\ifx\csname l@#1\endcsname\relax
\typeout{** WARNING: IEEEtran.bst: No hyphenation pattern has been}%
\typeout{** loaded for the language `#1'. Using the pattern for}%
\typeout{** the default language instead.}%
\else
\language=\csname l@#1\endcsname
\fi
#2}}
\providecommand{\BIBdecl}{\relax}
\BIBdecl

\bibitem{fortune-smart-home}
``{The global smart home market size was valued at \$80.21 billion in 2022 \&
  is projected to grow from \$93.98 billion in 2023 to \$338.28 billion by
  2030.}''
  \url{https://www.fortunebusinessinsights.com/industry-reports/smart-home-market-101900},
  2024.

\bibitem{smart-home-market}
``{With 21.1\% CAGR, Smart Home Market Worth USD 380.52 Billion in 2028},''
  \url{https://www.globenewswire.com/en/news-release/2022/06/28/2470249/0/en/With-21-1-CAGR-Smart-Home-Market-Worth-USD-380-52-Billion-in-2028.html},
  2022.

\bibitem{smart_things}
\BIBentryALTinterwordspacing
SmartThings, \emph{SmartThings}, 2024. [Online]. Available:
  \url{https://www.smartthings.com/}
\BIBentrySTDinterwordspacing

\bibitem{openhab_vs_ha}
\BIBentryALTinterwordspacing
P.~Lounge, \emph{OpenHAB Vs Home Assistant: Detailed Comparison By An Expert in
  2022}, Retrieved September 12, 2022. [Online]. Available:
  \url{https://purdylounge.com/openhab-vs-home-assistant/}
\BIBentrySTDinterwordspacing

\bibitem{if-start-again}
``{If starting again what smart home platform would people go with ?}''
  \url{https://www.reddit.com/r/smarthome/comments/16sk3qg/if_starting_again_what_smart_home_platform_would/},
  2023.

\bibitem{the-best}
``{The best home automation systems: Compare SmartThings, Apple HomeKit, Amazon
  Alexa, and more},''
  \url{https://www.zdnet.com/home-and-office/smart-home/best-home-automation-system/},
  2023.

\bibitem{how-popular}
``{How Popular Is Home Assistant Compared To The Competition?}''
  \url{https://whatsmarthome.com/how-popular-is-home-assistant/}, 2023.

\bibitem{haa}
``{Home Assistant Analytics},'' \url{https://analytics.home-assistant.io},
  2024.

\bibitem{yaml}
``{YAML},'' \url{https://yaml.org}, 2024.

\bibitem{yaml-example}
``{Automation Conditions - Home Assistant},''
  \url{https://www.home-assistant.io/docs/automation/condition/}, 2024.

\bibitem{hac}
``{Home Assistant Community},'' \url{https://community.home-assistant.io},
  2024.

\bibitem{tags}
``{Tags - Home Assistant Community},''
  \url{https://community.home-assistant.io/tags}, 2024.

\bibitem{restart-automations}
``{Strategies for dealing with restarts in automations},''
  \url{https://community.home-assistant.io/t/strategies-for-dealing-with-restarts-in-automations/400922},
  2022.

\bibitem{house-thermostat}
``{Automation - house thermostat based on phone location - not working},''
  \url{https://community.home-assistant.io/t/automation-house-thermostat-based-on-phone-location-not-working/401942},
  2022.

\bibitem{going-crazy}
``{Time Before and After condition, I'm going crazy},''
  \url{https://community.home-assistant.io/t/time-before-and-after-condition-i-m-going-crazy/416170},
  2022.

\bibitem{ifttt}
``{IFTTT - Automation for business and home},'' \url{https://ifttt.com}, 2024.

\bibitem{trigger}
``{Automation Trigger - Home Assistant},''
  \url{https://www.home-assistant.io/docs/automation/trigger/}, 2024.

\bibitem{action}
``{Automation actions - Home Assistant},''
  \url{https://www.home-assistant.io/docs/automation/action/}, 2024.

\bibitem{Jinja}
``Jinja,'' \url{https://palletsprojects.com/p/jinja/}, 2022.

\bibitem{github}
``{GitHub},'' \url{https://github.com}, 2024.

\bibitem{script-syntax}
``{Script Syntax - Home Assistant},''
  \url{https://www.home-assistant.io/docs/scripts/}, 2022.

\bibitem{on}
``{Boolean triggering script},''
  \url{https://community.home-assistant.io/t/boolean-triggering-script/404465},
  2022.

\bibitem{not-to}
``{Condition on trigger.entity\_id},''
  \url{https://community.home-assistant.io/t/condition-on-trigger-entity-id/442838},
  2022.

\bibitem{template-in-trigger-question}
``{Cant use template in for field of trigger automation},''
  \url{https://community.home-assistant.io/t/cant-use-template-in-for-field-of-trigger-automation/400475},
  2022.

\bibitem{impl-state}
``{Putting information from sensor into mobile notification},''
  \url{https://community.home-assistant.io/t/putting-information-from-sensor-into-mobile-notification/264000},
  2022.

\bibitem{debug-state}
``{Notify - How to display sensor value in message},''
  \url{https://community.home-assistant.io/t/notify-how-to-display-sensor-value-in-message/376646},
  2022.

\bibitem{wrong-predicates}
``{Lights on only after sunset and before sunrise},''
  url{https://community.home-assistant.io/t/lights-on-only-after-sunset-and-before-sunrise/307761},
  2022.

\bibitem{wrong-service-data}
``{My automation generates an error message},''
  \url{https://community.home-assistant.io/t/my-automation-generates-an-error-message/434970},
  2022.

\bibitem{quote}
``{Set counter value equals to other counter in automation},''
  \url{https://community.home-assistant.io/t/set-counter-value-equals-to-other-counter-in-automation/442485},
  2022.

\bibitem{indent}
``{Automation door alarm use template},''
  \url{https://community.home-assistant.io/t/automation-door-alarm-use-template/439744},
  2022.

\bibitem{exceptions}
``{Exceptions on automation},''
  \url{https://community.home-assistant.io/t/exceptions-on-automation/409661},
  2022.

\bibitem{automation-reference}
``{Automation - reference condition template variable in action
  notification},''
  \url{https://community.home-assistant.io/t/automation-reference-condition-template-variable-in-action-notification/339858},
  2021.

\bibitem{automation-action}
``{Automation Action Templating},''
  \url{https://community.home-assistant.io/t/automation-action-templating/409680},
  2022.

\bibitem{codeBeautify_yaml}
\BIBentryALTinterwordspacing
C.~2023, \emph{CodeBeautify - YAML Validator}, Accessed October 11, 2023.
  [Online]. Available: \url{https://codebeautify.org/yaml-validator}
\BIBentrySTDinterwordspacing

\bibitem{yaml_lint}
\BIBentryALTinterwordspacing
Y.~Lint, \emph{YAML Lint}, Accessed October 11, 2023. [Online]. Available:
  \url{https://www.yamllint.com/}
\BIBentrySTDinterwordspacing

\bibitem{yaml_checker}
\BIBentryALTinterwordspacing
Y.~Checker, \emph{YAML Checker}, Accessed October 11, 2023. [Online].
  Available: \url{https://yamlchecker.com/}
\BIBentrySTDinterwordspacing

\bibitem{json_formatter_yaml}
\BIBentryALTinterwordspacing
J.~Formatter, \emph{JSON Formatter - YAML Validator}, Accessed October 11,
  2023. [Online]. Available: \url{https://jsonformatter.org/yaml-validator}
\BIBentrySTDinterwordspacing

\bibitem{online_yaml_tools}
\BIBentryALTinterwordspacing
O.~Y. Tools, \emph{Online YAML Tools}, Accessed October 11, 2023. [Online].
  Available: \url{https://onlineyamltools.com/validate-yaml}
\BIBentrySTDinterwordspacing

\bibitem{ha}
``{Home Assistant},'' \url{https://www.home-assistant.io}, 2022.

\bibitem{fernandes2016security}
E.~Fernandes, J.~Jung, and A.~Prakash, ``Security analysis of emerging smart
  home applications,'' in \emph{2016 IEEE symposium on security and privacy
  (SP)}.\hskip 1em plus 0.5em minus 0.4em\relax IEEE, 2016, pp. 636--654.

\bibitem{He2019}
W.~He, J.~Martinez, R.~Padhi, L.~Zhang, and B.~Ur, ``When smart devices are
  stupid: Negative experiences using home smart devices,'' in \emph{2019 IEEE
  Security and Privacy Workshops (SPW)}, 2019, pp. 150--155.

\bibitem{alrawi2019sok}
O.~Alrawi, C.~Lever, M.~Antonakakis, and F.~Monrose, ``Sok: Security evaluation
  of home-based iot deployments,'' in \emph{2019 IEEE symposium on security and
  privacy (sp)}.\hskip 1em plus 0.5em minus 0.4em\relax IEEE, 2019, pp.
  1362--1380.

\bibitem{Brackenbury2019}
\BIBentryALTinterwordspacing
W.~Brackenbury, A.~Deora, J.~Ritchey, J.~Vallee, W.~He, G.~Wang, M.~L. Littman,
  and B.~Ur, ``How users interpret bugs in trigger-action programming,'' in
  \emph{Proceedings of the 2019 CHI Conference on Human Factors in Computing
  Systems}, ser. CHI '19.\hskip 1em plus 0.5em minus 0.4em\relax New York, NY,
  USA: Association for Computing Machinery, 2019, pp. 1--12. [Online].
  Available: \url{https://doi.org/10.1145/3290605.3300782}
\BIBentrySTDinterwordspacing

\bibitem{Zhou2021}
W.~Zhou, C.~Cao, D.~Huo, K.~Cheng, L.~Zhang, L.~Guan, T.~Liu, Y.~Jia, Y.~Zheng,
  Y.~Zhang, L.~Sun, Y.~Wang, and P.~Liu, ``Reviewing iot security via logic
  bugs in iot platforms and systems,'' \emph{IEEE Internet of Things Journal},
  vol.~8, no.~14, pp. 11\,621--11\,639, 2021.

\bibitem{makhshari2021iot}
A.~Makhshari and A.~Mesbah, ``Iot bugs and development challenges,'' in
  \emph{2021 IEEE/ACM 43rd International Conference on Software Engineering
  (ICSE)}.\hskip 1em plus 0.5em minus 0.4em\relax IEEE, 2021, pp. 460--472.

\bibitem{wang2022understanding}
T.~Wang, K.~Zhang, W.~Chen, W.~Dou, J.~Zhu, J.~Wei, and T.~Huang,
  ``Understanding device integration bugs in smart home system,'' in
  \emph{Proceedings of the 31st ACM SIGSOFT International Symposium on Software
  Testing and Analysis}, 2022, pp. 429--441.

\bibitem{liang2016understanding}
H.~Liang, Q.~Zhao, Y.~Wang, and H.~Liu, ``Understanding and detecting
  performance and security bugs in iot oses,'' in \emph{2016 17th IEEE/ACIS
  International Conference on Software Engineering, Artificial Intelligence,
  Networking and Parallel/Distributed Computing (SNPD)}.\hskip 1em plus 0.5em
  minus 0.4em\relax IEEE, 2016, pp. 413--418.

\bibitem{celik2018soteria}
Z.~B. Celik, P.~McDaniel, and G.~Tan, ``Soteria: Automated $\{$IoT$\}$ safety
  and security analysis,'' in \emph{2018 USENIX Annual Technical Conference
  (USENIX ATC 18)}, 2018, pp. 147--158.

\bibitem{trimananda2020understanding}
R.~Trimananda, S.~A.~H. Aqajari, J.~Chuang, B.~Demsky, G.~H. Xu, and S.~Lu,
  ``Understanding and automatically detecting conflicting interactions between
  smart home iot applications,'' in \emph{Proceedings of the 28th ACM Joint
  Meeting on European Software Engineering Conference and Symposium on the
  Foundations of Software Engineering}, 2020, pp. 1215--1227.

\bibitem{Li2020}
\BIBentryALTinterwordspacing
X.~Li, L.~Zhang, and X.~Shen, ``Diac: An inter-app conflicts detector for open
  iot systems,'' \emph{ACM Trans. Embed. Comput. Syst.}, vol.~19, no.~6, oct
  2020. [Online]. Available: \url{https://doi.org/10.1145/3391895}
\BIBentrySTDinterwordspacing

\bibitem{Xiao2020}
F.~Xiao, L.-T. Sha, Z.-P. Yuan, and R.-C. Wang, ``Vulhunter: A discovery for
  unknown bugs based on analysis for known patches in industry internet of
  things,'' \emph{IEEE Transactions on Emerging Topics in Computing}, vol.~8,
  no.~2, pp. 267--279, 2020.

\bibitem{Fu2021}
\BIBentryALTinterwordspacing
L.~Fu, S.~Ji, K.~Lu, P.~Liu, X.~Zhang, Y.~Duan, Z.~Zhang, W.~Chen, and Y.~Wu,
  ``Cpscan: Detecting bugs caused by code pruning in iot kernels,'' ser. CCS
  '21.\hskip 1em plus 0.5em minus 0.4em\relax New York, NY, USA: Association
  for Computing Machinery, 2021, pp. 794--810. [Online]. Available:
  \url{https://doi.org/10.1145/3460120.3484738}
\BIBentrySTDinterwordspacing

\bibitem{Huang2023}
\BIBentryALTinterwordspacing
B.~Huang, D.~Chaki, A.~Bouguettaya, and K.-Y. Lam, ``A survey on conflict
  detection in iot-based smart homes,'' \emph{ACM Comput. Surv.}, vol.~56,
  no.~5, nov 2023. [Online]. Available: \url{https://doi.org/10.1145/3629517}
\BIBentrySTDinterwordspacing

\end{thebibliography}

\end{document}